\documentclass{article}

\usepackage{graphicx}

\begin{document}

\title{Probability and QCD}
\date{September 28, 2017}
\author{G. Quznetsov \\ 
mailto: quznetso@gmail.com}
\maketitle

\abstract{The probabilities of point events in space 3 + 1 obey an equation of Dirac type.

Masses, moments, energies, spins, etc. are the parameters of the probability distribution of such events.

The terms and equations of quark-gluon theories turn out theoretically probabilistic terms and theorems.
Confinement and asymptotic freedom are explained by behaviour of such probabolities. And here we have the probabilistic foundations of the theory of gravitation.

Knowledge of the elements of linear algebra and differential calculus is sufficient to understand the content of this article}

\tableofcontents

\section{Introduction}
\label{sec:intro}

	In the study of the logical foundations of probability theory \cite{Q1}, I found that the terms and equations of the fundamental theoretical physics 
represent terms and theorems of the classical probability theory, more precisely, of that part of this theory, which considers the probability of 
dot events in the 3 + 1 space-time.

	In particular, all Standard Model's formulas (higgs ones except) turn out theorems of such probability theory. And the masses, moments, 
energies, spins, etc. turn out of parameters of probability distributions such events. The terms and the equations of the electroweak and of the 
quark-gluon theories turn out the theoretical-probabilistic terms and theorems. Here the relation of a neutrino to his lepton becomes clear, 
the W and Z bosons masses turn out dynamic ones, the cause of the asymmetry between particles and antiparticles is the impossibility of the birth 
of single antiparticles. In addition, phenomena such as confinement and asymptotic freedom receive their probabilistic explanation. And here we have 
the logical foundations of the gravity theory with phenomena dark energy and dark matter.

The proposed article contains initial concepts and the results of the development of these ideas.

\section{Propagation of probability}

Let us consider dot events in the 3+1 space-time with coordinates:

\begin{eqnarray*}
\mathbf{x} &:&=\left( x_1,x_2,x_3\right) \mbox{,} \\
\underline{x} &:&=\left( x_0,\mathbf{x}\right) \mbox{,} \\
\int d^{3+1}\underline{x} &:&=\int dx_0\int dx_1\int dx_2\int dx_3\mbox{,} \\
\int d^3\mathbf{y} &:&=\int dy_1\int dy_2\int dy_3\mbox{,} \\
t &:&=\frac{x_0}{\mathrm{c}}\mbox{.}
\end{eqnarray*}

Sentence of type: $\ll $Event $\mathsf{A}$ occurs in point $\underline{x}\gg 
$ will be written the followig way: $\ll \mathsf{A}\left( \underline{x}%
\right) \gg $\textquotedblright .

Events of type $^{\circ }\ll \mathsf{A}\left( \underline{x}\right) \gg $ are
called \textit{dot events.} All dot events and all events received from dot
events by operations of addition, multiplication and addition, are \textit{%
physical events}.

$\mathsf{A}(D)$ means: $\left( \mathsf{A}\left( \underline{x}\right) \wedge
^{\circ }\ll \left( \underline{x}\right) \in D\gg \right) $.

Let $\mathrm{P}$ be the probability function.

Let $\left\langle X_{\mathsf{A,}0},X_{\mathsf{A,}1},X_{\mathsf{A,}2},X_{%
\mathsf{A,}3}\right\rangle $ be random coordinates of event $\mathsf{A}$.

Let $F_{\mathsf{A}}$ be a \textit{Cumulative Distribution Function} i.e.:

\[
F_{\mathsf{A}}\left( x_{0},x_{1},x_{2},x_{3}\right) =\mathrm{P}\left( \left(
X_{\mathsf{A,}0}<x_{0}\right) \wedge \left( X_{\mathsf{A,}1}<x_{1}\right)
\wedge \left( X_{\mathsf{A,}2}<x_{2}\right) \wedge \left( X_{\mathsf{A,}%
3}<x_{3}\right) \right) \mbox{.}
\]

If

\begin{eqnarray*}
j_{0} &:&=\frac{\partial ^{3}F}{\partial x_{1}\partial x_{2}\partial x_{3}}%
\mbox{,} \\
j_{1} &:&=-\frac{\partial ^{3}F}{\partial x_{0}\partial x_{2}\partial x_{3}}%
\mbox{,} \\
j_{2} &:&=-\frac{\partial ^{3}F}{\partial x_{0}\partial x_{1}\partial x_{3}}%
\mbox{,} \\
j_{3} &:&=\frac{\partial ^{3}F}{\partial x_{0}\partial x_{1}\partial x_{2}}
\end{eqnarray*}

then $\left\langle j_{0},j_{1},j_{2},j_{3}\right\rangle $ is \textit{a
probability current vector of event}.

If $\rho :=j_{0}/\mathrm{c}$ then $\rho $ is \textit{a a probability density
function}.

If $\ \mathbf{u}_{\mathsf{A}}:=\mathbf{j}_{\mathsf{A}}/\rho _{\mathsf{A}}$
then vector $\mathbf{u}_{\mathsf{A}}$ is \textit{a velocity of the
probability} of ${\mathsf{A}}$ propagation.

(for example for $u_{2}$: 
\[
u_{2}=\frac{j_{2}}{\rho }=\frac{\left( -\frac{\partial ^{3}F}{\partial
x_{0}\partial x_{1}\partial x_{3}}\right) c}{\left( \frac{\partial ^{3}F}{%
\partial x_{1}\partial x_{2}\partial x_{3}}\right) }=\left( -\frac{\Delta
_{013}F}{\Delta _{123}F}\frac{\Delta x_{2}}{\Delta x_{0}}\right) c 
\]%
)

Probability, for which $u_{1}^{2}+u_{2}^{2}+u_{3}^{2}\leq \mathrm{c}$ are
called \textit{traceable probability}.

Denote:

\[
1_2:=\left[ 
\begin{array}{cc}
1 & 0 \\ 
0 & 1%
\end{array}
\right] \mbox{, }0_2:=\left[ 
\begin{array}{cc}
0 & 0 \\ 
0 & 0%
\end{array}
\right] \mbox{, }\beta ^{[0]}:=-\left[ 
\begin{array}{cc}
1_2 & 0_2 \\ 
0_2 & 1_2%
\end{array}
\right] =-1_4\mbox{,} 
\]

the Pauli matrices

\[
\sigma _1=\left[ 
\begin{array}{cc}
0 & 1 \\ 
1 & 0%
\end{array}
\right] \mbox{, }\sigma _2=\left[ 
\begin{array}{cc}
0 & -\mathrm{i} \\ 
\mathrm{i} & 0%
\end{array}
\right] \mbox{, }\sigma _3=\left[ 
\begin{array}{cc}
1 & 0 \\ 
0 & -1%
\end{array}
\right] \mbox{.} 
\]

A set $\widetilde{C}$ of complex $n\times n$ matrices is called \textit{a
Clifford set} \textit{of rank $n$} \cite{Md} if the
following conditions are fulfilled:

if $\alpha _k\in \widetilde{C}$ and $\alpha _r\in \widetilde{C}$ then $%
\alpha _k\alpha _r+\alpha _r\alpha _k=2\delta _{k,r}$;

if $\alpha _k\alpha _r+\alpha _r\alpha _k=2\delta _{k,r}$ for all elements $%
\alpha _r$ of set $\widetilde{C}$ then $\alpha _k\in \widetilde{C}$.

If $n=4$ then a Clifford set either contains $3$ matrices (\textit{a
Clifford triplet}) or contains $5$ matrices (\textit{a Clifford pentad}).

Here exist only six Clifford pentads \cite{Md}: one \textit{light pentad} $%
\beta $: 
\begin{equation}
\begin{array}{c}
\beta ^{\left[ 1\right] }:=\left[ 
\begin{array}{cc}
\sigma _{1} & 0_{2} \\ 
0_{2} & -\sigma _{1}%
\end{array}%
\right] \mbox{, }\beta ^{\left[ 2\right] }:=\left[ 
\begin{array}{cc}
\sigma _{2} & 0_{2} \\ 
0_{2} & -\sigma _{2}%
\end{array}%
\right] \mbox{, } \\ 
\beta ^{\left[ 3\right] }:=\left[ 
\begin{array}{cc}
\sigma _{3} & 0_{2} \\ 
0_{2} & -\sigma _{3}%
\end{array}%
\right] \mbox{,}%
\end{array}
\label{lghr}
\end{equation}%
\begin{equation}
\gamma ^{\left[ 0\right] }:=\left[ 
\begin{array}{cc}
0_{2} & 1_{2} \\ 
1_{2} & 0_{2}%
\end{array}%
\right] \mbox{, }  \label{lghr1}
\end{equation}%
\begin{equation}
\beta ^{\left[ 4\right] }:=\mathrm{i}\cdot \left[ 
\begin{array}{cc}
0_{2} & 1_{2} \\ 
-1_{2} & 0_{2}%
\end{array}%
\right] \mbox{;}  \label{lghr2}
\end{equation}

three \textit{chromatic} pentads:

\textit{the red pentad} $\zeta $: 
\begin{equation}
\zeta ^{[1]}=\left[ 
\begin{array}{cc}
-\sigma _1 & 0_2 \\ 
0_2 & \sigma _1%
\end{array}
\right] ,\zeta ^{[2]}=\left[ 
\begin{array}{cc}
\sigma _2 & 0_2 \\ 
0_2 & \sigma _2%
\end{array}
\right] ,\zeta ^{[3]}=\left[ 
\begin{array}{cc}
-\sigma _3 & 0_2 \\ 
0_2 & -\sigma _3%
\end{array}
\right] \mbox{,}  \label{red}
\end{equation}
\begin{equation}
\gamma _\zeta ^{[0]}=\left[ 
\begin{array}{cc}
0_2 & -\sigma _1 \\ 
-\sigma _1 & 0_2%
\end{array}
\right] \mbox{, }\zeta ^{[4]}=\mathrm{i} \left[ 
\begin{array}{cc}
0_2 & \sigma _1 \\ 
-\sigma _1 & 0_2%
\end{array}
\right] \mbox{;}  \label{redM}
\end{equation}

\textit{the green pentad} $\eta $: 
\begin{equation}
\eta ^{[1]}=\left[ 
\begin{array}{cc}
-\sigma _1 & 0_2 \\ 
0_2 & -\sigma _1%
\end{array}
\right] ,\eta ^{[2]}=\left[ 
\begin{array}{cc}
-\sigma _2 & 0_2 \\ 
0_2 & \sigma _2%
\end{array}
\right] ,\eta ^{[3]}=\left[ 
\begin{array}{cc}
\sigma _3 & 0_2 \\ 
0_2 & \sigma _3%
\end{array}
\right] \mbox{,}  \label{green}
\end{equation}
\begin{equation}
\gamma _\eta ^{[0]}=\left[ 
\begin{array}{cc}
0_2 & -\sigma _2 \\ 
-\sigma _2 & 0_2%
\end{array}
\right] \mbox{, }\eta ^{[4]}=\mathrm{i} \left[ 
\begin{array}{cc}
0_2 & \sigma _2 \\ 
-\sigma _2 & 0_2%
\end{array}
\right] \mbox{;}  \label{greenM}
\end{equation}

\textit{the blue pentad} $\theta $: 
\begin{equation}
\theta ^{[1]}=\left[ 
\begin{array}{cc}
\sigma _1 & 0_2 \\ 
0_2 & \sigma _1%
\end{array}
\right] ,\theta ^{[2]}=\left[ 
\begin{array}{cc}
-\sigma _2 & 0_2 \\ 
0_2 & -\sigma _2%
\end{array}
\right] ,\theta ^{[3]}=\left[ 
\begin{array}{cc}
-\sigma _3 & 0_2 \\ 
0_2 & \sigma _3%
\end{array}
\right] \mbox{,}  \label{blue}
\end{equation}
\begin{equation}
\gamma _\theta ^{[0]}=\left[ 
\begin{array}{cc}
0_2 & -\sigma _3 \\ 
-\sigma _3 & 0_2%
\end{array}
\right] ,\theta ^{[4]}=\mathrm{i} \left[ 
\begin{array}{cc}
0_2 & \sigma _3 \\ 
-\sigma _3 & 0_2%
\end{array}
\right] \mbox{;}  \label{blueM}
\end{equation}

two \textit{gustatory} pentads:

\textit{the sweet pentad} $\underline{\Delta }$: 
\[
\underline{\Delta }^{[1]}=\left[ 
\begin{array}{cc}
0_2 & -\sigma _1 \\ 
-\sigma _1 & 0_2%
\end{array}
\right] ,\underline{\Delta }^{[2]}=\left[ 
\begin{array}{cc}
0_2 & -\sigma _2 \\ 
-\sigma _2 & 0_2%
\end{array}
\right] ,\underline{\Delta }^{[3]}=\left[ 
\begin{array}{cc}
0_2 & -\sigma _3 \\ 
-\sigma _3 & 0_2%
\end{array}
\right] , 
\]
\[
\underline{\Delta }^{[0]}=\left[ 
\begin{array}{cc}
-1_2 & 0_2 \\ 
0_2 & 1_2%
\end{array}
\right] ,\underline{\Delta }^{[4]}=\mathrm{i} \left[ 
\begin{array}{cc}
0_2 & 1_2 \\ 
-1_2 & 0_2%
\end{array}
\right] ; 
\]

\textit{the bitter pentad} $\underline{\Gamma}$: 
\[
\underline{\Gamma }^{[1]}=\mathrm{i} \left[ 
\begin{array}{cc}
0_2 & -\sigma _1 \\ 
\sigma _1 & 0_2%
\end{array}
\right] ,\underline{\Gamma }^{[2]}=\mathrm{i} \left[ 
\begin{array}{cc}
0_2 & -\sigma _2 \\ 
\sigma _2 & 0_2%
\end{array}
\right] ,\underline{\Gamma }^{[3]}=\mathrm{i} \left[ 
\begin{array}{cc}
0_2 & -\sigma _3 \\ 
\sigma _3 & 0_2%
\end{array}
\right] , 
\]
\[
\underline{\Gamma }^{[0]}=\left[ 
\begin{array}{cc}
-1_2 & 0_2 \\ 
0_2 & 1_2%
\end{array}
\right] ,\underline{\Gamma }^{[4]}=\left[ 
\begin{array}{cc}
0_2 & 1_2 \\ 
1_2 & 0_2%
\end{array}
\right] \mbox{.} 
\]

Further we do not consider gustatory pentads since these pentads are not
used yet in the contemporary physics.

Let $\kappa :=\sum\limits_{s=0}^{3}\beta ^{\lbrack s]}x_{s}$.

If

\begin{equation}
U_{0,1}\left( \sigma \right) :=\left[ 
\begin{array}{cccc}
\cosh \sigma & -\sinh \sigma & 0 & 0 \\ 
-\sinh \sigma & \cosh \sigma & 0 & 0 \\ 
0 & 0 & \cosh \sigma & \sinh \sigma \\ 
0 & 0 & \sinh \sigma & \cosh \sigma%
\end{array}%
\right]  \label{u01}
\end{equation}

then

\begin{eqnarray*}
U_{0,1}^{\dagger }\left( \sigma \right) \kappa U_{0,1}\left( \sigma \right) =
\\
\beta ^{\lbrack 0]}\left( x_{0}\cosh 2\sigma +x_{1}\sinh 2\sigma \right) \\
+\beta ^{\lbrack 1]}\left( x_{1}\cosh 2\sigma +x_{0}\sinh 2\sigma \right) \\
+\beta ^{\lbrack 2]}x_{2}+\beta ^{\lbrack 3]}x_{3} \mbox{.}
\end{eqnarray*}

Hence, $U_{0,1}$ makes the Lorentz transformaton $\left\langle
x_{0},x_{1}\right\rangle $:

\begin{eqnarray*}
x_{0} &\rightarrow &x_{0}^{\prime }:=x_{0}\cosh 2\sigma +x_{1}\sinh 2\sigma %
\mbox{,} \\
x_{1} &\rightarrow &x_{1}^{\prime }:=x_{1}\cosh 2\sigma +x_{0}\sinh 2\sigma %
\mbox{,} \\
x_{2} &\rightarrow &x_{2}^{\prime }:=x_{2} \mbox{,} \\
x_{3} &\rightarrow &x_{3}^{\prime }:=x_{3} \mbox{.}
\end{eqnarray*}

Similarly,

\begin{equation}
U_{0,2}\left( \phi \right) :=\left[ 
\begin{array}{cccc}
\cosh \phi & \mathrm{i}\sinh \phi & 0 & 0 \\ 
-\mathrm{i}\sinh \phi & \cosh \phi & 0 & 0 \\ 
0 & 0 & \cosh \phi & -\mathrm{i}\sinh \phi \\ 
0 & 0 & \mathrm{i}\sinh \phi & \cosh \phi%
\end{array}%
\right]  \label{u02}
\end{equation}

makes the Lorentz transformaton $\left\langle x_{0},x_{2}\right\rangle $ and

\begin{equation}
U_{0,3}\left( \iota \right) :=\left[ 
\begin{array}{cccc}
e^{\iota } & 0 & 0 & 0 \\ 
0 & e^{-\iota } & 0 & 0 \\ 
0 & 0 & e^{-\iota } & 0 \\ 
0 & 0 & 0 & e^{\iota }%
\end{array}%
\right]  \label{u03}
\end{equation}

makes the Lorentz transformaton $\left\langle x_{0},x_{3}\right\rangle $.

If

\begin{equation}
U_{1,3}\left( \vartheta \right) :=\left[ 
\begin{array}{cccc}
\cos \vartheta & \sin \vartheta & 0 & 0 \\ 
-\sin \vartheta & \cos \vartheta & 0 & 0 \\ 
0 & 0 & \cos \vartheta & \sin \vartheta \\ 
0 & 0 & -\sin \vartheta & \cos \vartheta%
\end{array}%
\right]  \label{u13}
\end{equation}

then $U_{1,3}\left( \vartheta \right) $ makes the cartesian turn $%
\left\langle x_{1},x_{3}\right\rangle :$

\begin{eqnarray*}
&&U_{1,3}^{\dagger }\left( \vartheta \right) \kappa U_{1,3}\left( \vartheta
\right) \\
&=&\beta ^{\lbrack 0]}x_{0} \\
&&+\beta ^{\lbrack 1]}\left( x_{1}\cos 2\vartheta +x_{3}\sin 2\vartheta
\right) \\
&&+\beta ^{\lbrack 2]}x_{2} \\
&&+\beta ^{\lbrack 3]}\left( x_{3}\cos 2\vartheta -x_{1}\sin 2\vartheta
\right)
\end{eqnarray*}

Similarly,

\begin{equation}
U_{1,2}\left( \varsigma \right) :=\left[ 
\begin{array}{cccc}
e^{-\mathrm{i}\varsigma } & 0 & 0 & 0 \\ 
0 & e^{\mathrm{i}\varsigma } & 0 & 0 \\ 
0 & 0 & e^{-\mathrm{i}\varsigma } & 0 \\ 
0 & 0 & 0 & e^{\mathrm{i}\varsigma }%
\end{array}%
\right]  \label{u12}
\end{equation}

makes the cartesian turn $\left\langle x_{1},x_{2}\right\rangle $ and

\begin{equation}
U_{2,3}\left( \alpha \right) :=\left[ 
\begin{array}{cccc}
\cos \alpha & \mathrm{i}\sin \alpha & 0 & 0 \\ 
\mathrm{i}\sin \alpha & \cos \alpha & 0 & 0 \\ 
0 & 0 & \cos \alpha & \mathrm{i}\sin \alpha \\ 
0 & 0 & \mathrm{i}\sin \alpha & \cos \alpha%
\end{array}%
\right]  \label{u23}
\end{equation}

makes the cartesian turn $\left\langle x_{2},x_{3}\right\rangle $.

Let us consider the following set of four real equations with eight real
unknowns: $b^2$ with $b>0$, ${\alpha }$, ${\beta }$, ${\chi }$, ${\theta }$, 
${\gamma }$, ${\upsilon }$, ${\lambda }$:

\begin{equation}
\left\{ 
\begin{array}{c}
b^2=\rho \mbox{,} \\ 
b^2\left( \cos ^2\left( {\alpha }\right) \sin \left( 2{\beta }\right) \cos
\left( {\theta }-{\gamma }\right) -\sin ^2\left( {\alpha }\right) \sin
\left( 2{\chi }\right) \cos \left( {\upsilon }-{\lambda }\right) \right) =-%
\frac{j_{1}}{\mathrm{c}}\mbox{,} \\ 
b^2\left( \cos ^2\left( {\alpha }\right) \sin \left( 2{\beta }\right) \sin
\left( {\theta }-{\gamma }\right) -\sin ^2\left( {\alpha }\right) \sin
\left( 2{\chi }\right) \sin \left( {\upsilon }-{\lambda }\right) \right) =-%
\frac{j_{2}}{\mathrm{c}}\mbox{,} \\ 
b^2\left( \cos ^2\left( {\alpha }\right) \cos \left( 2{\beta }\right) -\sin
^2\left( {\alpha }\right) \cos \left( 2{\chi }\right) \right) =-\frac{j_{3}}{%
\mathrm{c}}\mbox{.}%
\end{array}
\right|  \label{abc}
\end{equation}

This set has solutions for any traceable $\rho $ and $j_{\mathcal{A},k}$.
For example one of these solutions is the following:

1. A value of $b^2$ obtain from first equation.

2. Since

\[
u_{k}=\frac{j_{k}}{\rho } 
\]

then 
\[
\left\{ 
\begin{array}{c}
\cos ^2\left( {\alpha }\right) \sin \left( 2{\beta }\right) \cos \left( {%
\theta }-{\gamma }\right) -\sin ^2\left( {\alpha }\right) \sin \left( 2{\chi 
}\right) \cos \left( {\upsilon }-{\lambda }\right) =-\frac{u_{1}}{\mathrm{c}}%
\mbox{,} \\ 
\cos ^2\left( {\alpha }\right) \sin \left( 2{\beta }\right) \sin \left( {%
\theta }-{\gamma }\right) -\sin ^2\left( {\alpha }\right) \sin \left( 2{\chi 
}\right) \sin \left( {\upsilon }-{\lambda }\right) =-\frac{u_{2}}{\mathrm{c}}%
\mbox{,} \\ 
\cos ^2\left( {\alpha }\right) \cos \left( 2{\beta }\right) -\sin ^2\left( {%
\alpha }\right) \cos \left( 2{\chi }\right) =-\frac{u_{3}}{\mathrm{c}}%
\mbox{.}%
\end{array}
\right| 
\]

3. Let ${\beta =\chi }$.

In that case:

\[
\left\{ 
\begin{array}{c}
\left( \cos ^2\left( {\alpha }\right) \cos \left( {\theta }-{\gamma }\right)
-\sin ^2\left( {\alpha }\right) \cos \left( {\upsilon }-{\lambda }\right)
\right) \sin \left( 2{\beta }\right) =-\frac{u_{1}}{\mathrm{c}}\mbox{,} \\ 
\left( \cos ^2\left( {\alpha }\right) \sin \left( {\theta }-{\gamma }\right)
-\sin ^2\left( {\alpha }\right) \sin \left( {\upsilon }-{\lambda }\right)
\right) \sin \left( 2{\beta }\right) =-\frac{u_{2}}{\mathrm{c}}\mbox{,} \\ 
\left( \cos ^2\left( {\alpha }\right) -\sin ^2\left( {\alpha }\right)
\right) \cos \left( 2{\beta }\right) =-\frac{u_{3}}{\mathrm{c}}\mbox{.}%
\end{array}
\right| 
\]

4. Let $\left( {\theta }-{\gamma }\right) =\left( {\upsilon }-{\lambda }%
\right) $.

In that case:

\[
\left\{ 
\begin{array}{c}
\cos \left( 2{\alpha }\right) \cos \left( {\theta }-{\gamma }\right) \sin
\left( 2{\beta }\right) =-\frac{u_{1}}{\mathrm{c}}\mbox{,} \\ 
\cos \left( 2{\alpha }\right) \sin \left( {\theta }-{\gamma }\right) \sin
\left( 2{\beta }\right) =-\frac{u_{2}}{\mathrm{c}}\mbox{,} \\ 
\cos \left( 2{\alpha }\right) \cos \left( 2{\beta }\right) =-\frac{u_{3}}{%
\mathrm{c}}\mbox{.}%
\end{array}
\right| 
\]

5. Let us raise to the second power the first and the second equations:

\[
\left\{ 
\begin{array}{c}
\cos ^2\left( 2{\alpha }\right) \cos ^2\left( {\theta }-{\gamma }\right)
\sin ^2\left( 2{\beta }\right) =\left( -\frac{u_{1}}{\mathrm{c}}\right) ^2%
\mbox{,} \\ 
\cos ^2\left( 2{\alpha }\right) \sin ^2\left( {\theta }-{\gamma }\right)
\sin ^2\left( 2{\beta }\right) =\left( -\frac{u_{2}}{\mathrm{c}}\right) ^2%
\mbox{,} \\ 
\cos \left( 2{\alpha }\right) \cos \left( 2{\beta }\right) =-\frac{u_{3}}{%
\mathrm{c}}\mbox{.}%
\end{array}
\right| 
\]

and let us summat these two equations:

\[
\left\{ 
\begin{array}{c}
\sin ^2\left( 2{\beta }\right) \cos ^2\left( 2{\alpha }\right) \left( \cos
^2\left( {\theta }-{\gamma }\right) +\sin ^2\left( {\theta }-{\gamma }%
\right) \right) \\ 
=\left( -\frac{u_{1}}{\mathrm{c}}\right) ^2+\left( -\frac{u_{2}}{\mathrm{c}}%
\right) ^2\mbox{,} \\ 
\cos \left( 2{\alpha }\right) \cos \left( 2{\beta }\right) =-\frac{u_{3}}{%
\mathrm{c}}\mbox{.}%
\end{array}
\right| 
\]

Hence:

\[
\left\{ 
\begin{array}{c}
\sin ^2\left( 2{\beta }\right) \cos ^2\left( 2{\alpha }\right) =\left( -%
\frac{u_{1}}{\mathrm{c}}\right) ^2+\left( -\frac{u_{2}}{\mathrm{c}}\right)
^2 \mbox{,} \\ 
\cos \left( 2{\alpha }\right) \cos \left( 2{\beta }\right) =-\frac{u_{3}}{%
\mathrm{c}}\mbox{.}%
\end{array}
\right| 
\]

6. Let us raise to the second power the second equation and add this
equation to the previous one:

\[
\left\{ 
\begin{array}{c}
\sin ^2\left( 2{\beta }\right) \cos ^2\left( 2{\alpha }\right) =\left( -%
\frac{u_{1}}{\mathrm{c}}\right) ^2+\left( -\frac{u_{2}}{\mathrm{c}}\right) ^2%
\mbox{,} \\ 
\cos ^2\left( 2{\alpha }\right) \cos ^2\left( 2{\beta }\right) =\left( -%
\frac{u_{3}}{\mathrm{c}}\right) ^2%
\end{array}
\right| 
\]

\[
\left( \sin ^2\left( 2{\beta }\right) +\cos ^2\left( 2{\beta }\right)
\right) \cos ^2\left( 2{\alpha }\right) =\left( -\frac{u_{1}}{\mathrm{c}}%
\right) ^2+\left( -\frac{u_{2}}{\mathrm{c}}\right) ^2+\left( -\frac{u_{3}}{%
\mathrm{c}}\right) ^2\mbox{,} 
\]

\begin{equation}
\cos ^2\left( 2{\alpha }\right) =\left( -\frac{u_{1}}{\mathrm{c}}\right)
^2+\left( -\frac{u_{2}}{\mathrm{c}}\right) ^2+\left( -\frac{u_{3}}{\mathrm{c}%
}\right) ^2\mbox{,}  \label{less}
\end{equation}

We receive $\cos ^2\left( 2{\alpha }\right) $ (for a trackeable probabilities).

7. From

\[
\cos ^2\left( 2{\alpha }\right) \cos ^2\left( 2{\beta }\right) =\left( -%
\frac{u_{3}}{\mathrm{c}}\right) ^2 
\]

we receive $\cos ^2\left( 2{\beta }\right) $.

8. From

\[
\cos ^2\left( 2{\alpha }\right) \cos ^2\left( {\theta }-{\gamma }\right)
\sin ^2\left( 2{\beta }\right) =\left( -\frac{u_{1}}{\mathrm{c}}\right) ^2 
\]

we receive $\cos ^2\left( {\theta }-{\gamma }\right) $.

----------------------------------------------

If 
\begin{eqnarray}
&&\varphi _{1}:=b\exp \left( \mathrm{i}{\gamma }\right) \cos \left( {\beta }%
\right) \cos \left( {\alpha }\right) \mbox{,}  \nonumber \\
&&\varphi _{2}:=b\exp \left( \mathrm{i}{\theta }\right) \sin \left( {\beta }%
\right) \cos \left( {\alpha }\right) \mbox{,}  \nonumber \\
&&\varphi _{3}:=b\exp \left( \mathrm{i}{\lambda }\right) \cos \left( {\chi }%
\right) \sin \left( {\alpha }\right) \mbox{,}  \label{xxx55} \\
&&\varphi _{4}:=b\exp \left( \mathrm{i}{\upsilon }\right) \sin \left( {\chi }%
\right) \sin \left( {\alpha }\right)  \nonumber
\end{eqnarray}

then you can calculate that

\begin{eqnarray}
\rho &=&\sum_{s=1}^4\varphi _s^{*}\varphi _s\mbox{,}  \label{j} \\
\frac{j_{\alpha }}{\mathrm{c}} &=&-\sum_{k=1}^4\sum_{s=1}^4\varphi
_s^{*}\beta _{s,k}^{\left[ \alpha \right] }\varphi _k  \nonumber
\end{eqnarray}

If $\varphi ^{\prime }:=U_{0,2}\left( \phi \right) \varphi $ then

\[
\rho ^{\prime }=\varphi ^{\prime \dagger }\varphi ^{\prime }=\varphi
^{\dagger }U_{0,2}^{\dagger }\left( \phi \right) U_{0,2}\left( \phi \right)
\varphi =\rho \cosh 2\phi +\frac{j_{2}}{\mathrm{c}}\sinh 2\phi 
\]
and 
\[
\frac{j_{2}^{\prime }}{\mathrm{c}}=-\varphi _{s}^{\prime \dagger }\beta ^{%
\left[ 2\right] }\varphi _{k}^{\prime }=-\varphi ^{\dagger }U_{0,2}^{\dagger
}\left( \phi \right) \beta ^{\left[ 2\right] }U_{0,2}\left( \phi \right)
\varphi =\frac{j_{2}}{\mathrm{c}}\cosh 2\phi +\rho \sinh 2\phi \mbox{.} 
\]

Similarly $U_{0,1}$ and $U_{0,3}$ transform the 3+1 vector $\left\langle 
\mathrm{c}\rho ,\mathbf{j}\right\rangle $ by the Lorentz formulas and $%
U_{1,2}$, $U_{1,3}$, $U_{2,3}$ transform this vector by the cartesian
formulas.

Because

\[
\frac{\partial j_{0}}{\partial x_{0}}=\frac{\partial ^{4}F}{\partial
x_{0}\partial x_{1}\partial x_{2}\partial x_{3}}=-\frac{\partial j_{1}}{%
\partial x_{1}}=-\frac{\partial j_{2}}{\partial x_{2}}=\frac{\partial j_{3}}{%
\partial x_{3}} 
\]

then (\textit{Continuity equation}):

\begin{equation}
\frac{\partial \rho }{\partial x_{0}}+\frac{\partial j_{1}}{\partial x_{1}}+%
\frac{\partial j_{2}}{\partial x_{2}}+\frac{\partial j_{3}}{\partial x_{3}}=0
\label{NRZ}
\end{equation}

In that case:

\[
\frac{\partial \left( \varphi ^{\dagger }\varphi \right) }{\partial x_{0}}-%
\frac{\partial \left( \varphi ^{\dagger }\beta ^{\lbrack 1]}\varphi \right) 
}{\partial x_{1}}-\frac{\partial \left( \varphi ^{\dagger }\beta ^{\lbrack
2]}\varphi \right) }{\partial x_{2}}-\frac{\partial \left( \varphi ^{\dagger
}\beta ^{\lbrack 1]}\varphi \right) }{\partial x_{3}}=0 
\]

\begin{eqnarray*}
&&\frac{\partial \left( \varphi ^{\dagger }\right) }{\partial x_{0}}\varphi
+\varphi ^{\dagger }\frac{\partial \left( \varphi \right) }{\partial x_{0}}
\\
&&-\frac{\partial \left( \varphi ^{\dagger }\beta ^{\lbrack 1]}\right) }{%
\partial x_{1}}\varphi -\varphi ^{\dagger }\frac{\partial \left( \beta
^{\lbrack 1]}\varphi \right) }{\partial x_{1}} \\
&&-\frac{\partial \left( \varphi ^{\dagger }\beta ^{\lbrack 2]}\right) }{%
\partial x_{2}}\varphi -\varphi ^{\dagger }\frac{\partial \left( \beta
^{\lbrack 2]}\varphi \right) }{\partial x_{2}} \\
&&-\frac{\partial \left( \varphi ^{\dagger }\beta ^{\lbrack 3]}\right) }{%
\partial x_{3}}\varphi -\varphi ^{\dagger }\frac{\partial \left( \beta
^{\lbrack 3]}\varphi \right) }{\partial x_{3}} \\
&=&0
\end{eqnarray*}

\begin{eqnarray*}
&&\varphi ^{\dagger }\left( \frac{\partial }{\partial x_{0}}-\beta ^{\lbrack
1]}\frac{\partial }{\partial x_{1}}-\beta ^{\lbrack 2]}\frac{\partial }{%
\partial x_{2}}-\beta ^{\lbrack 3]}\frac{\partial }{\partial x_{3}}\right)
^{\dagger }\varphi \\
&&+\varphi ^{\dagger }\left( \frac{\partial }{\partial x_{0}}-\beta
^{\lbrack 1]}\frac{\partial }{\partial x_{1}}-\beta ^{\lbrack 2]}\frac{%
\partial }{\partial x_{2}}-\beta ^{\lbrack 3]}\frac{\partial }{\partial x_{3}%
}\right) \varphi \\
&=&0
\end{eqnarray*}

Let 
\begin{equation}
\widehat{Q}:=\frac{\partial }{x_{0}}-\sum\limits_{s=1}^{3}\beta ^{\lbrack s]}%
\frac{\partial }{x_{s}}  \label{Q}
\end{equation}

Hence,

\[
\varphi ^{\dagger }\left( \widehat{Q}^{\dagger }+\widehat{Q}\right) \varphi
=0 
\]

\begin{equation}
\widehat{Q}^{\dagger }=-\widehat{Q}  \label{H}
\end{equation}

Therefore, for every function $\varphi _{j}$ here exists an operator $%
Q_{j,k} $ such that a dependence of $\varphi _{j}$ on $t$ is described by
the following differential equations

\begin{equation}
\partial _t\varphi _j=\mathrm{c}\sum_{k=1}^4\left( \beta _{j,k}^{\left[ 1%
\right] }\partial _1+\beta _{j,k}^{\left[ 2\right] }\partial _2+\beta
_{j,k}^{\left[ 3\right] }\partial _3+Q_{j,k}\right) \varphi _k\mbox{.}
\label{ham}
\end{equation}

and $Q_{j,k}^{*}=-Q_{k,j}$.

A matrix form of formula (\ref{ham}) is the following: 
\begin{equation}
\partial _{t}\varphi =\mathrm{c}\left( \beta ^{\left[ 1\right] }\partial
_{1}+\beta ^{\left[ 2\right] }\partial _{2}+\beta ^{\left[ 3\right]
}\partial _{3}+\widehat{Q}\right) \varphi  \label{ham1}
\end{equation}

with 
\[
\varphi =\left[ 
\begin{array}{c}
\varphi _1 \\ 
\varphi _2 \\ 
\varphi _3 \\ 
\varphi _4%
\end{array}
\right] 
\]

and

\begin{equation}
\widehat{Q}=\left[ 
\begin{array}{cccc}
\mathrm{i}\vartheta _{1,1} & \mathrm{i}\vartheta _{1,2}-\varpi _{1,2} & 
\mathrm{i}\vartheta _{1,3}-\varpi _{1,3} & \mathrm{i}\vartheta _{1,4}-\varpi
_{1,4} \\ 
\mathrm{i}\vartheta _{1,2}+\varpi _{1,2} & \mathrm{i}\vartheta _{2,2} & 
\mathrm{i}\vartheta _{2,3}-\varpi _{2,3} & \mathrm{i}\vartheta _{2,4}-\varpi
_{2,4} \\ 
\mathrm{i}\vartheta _{1,3}+\varpi _{1,3} & \mathrm{i}\vartheta _{2,3}+\varpi
_{2,3} & \mathrm{i}\vartheta _{3,3} & \mathrm{i}\vartheta _{3,4}-\varpi
_{3,4} \\ 
\mathrm{i}\vartheta _{1,4}+\varpi _{1,4} & \mathrm{i}\vartheta _{2,4}+\varpi
_{2,4} & \mathrm{i}\vartheta _{3,4}+\varpi _{3,4} & \mathrm{i}\vartheta
_{4,4}%
\end{array}
\right]  \label{QR}
\end{equation}

with $\varpi _{s,k}=\mathrm{Re}\left( Q_{s,k}\right) $ and $\vartheta _{s,k}=%
\mathrm{Im}\left( Q_{s,k}\right) $. Matrix $\varphi $ is called \textit{a
state vector} of the event $\mathcal{A}$ probability. Let $\vartheta _{s,k}$
and $\varpi _{s,k}$ be terms of $\widehat{Q}$ (\ref{QR}) and let $\Theta
_{0} $, $\Theta _{3}$, $\Upsilon _{0}$ and $\Upsilon _{3}$ be a solution of
the following equations set: 
\[
\left\{ 
\begin{array}{c}
{-\Theta _{0}+\Theta _{3}-\Upsilon _{0}+\Upsilon _{3}}{=\vartheta _{1,1}}%
\mbox{;} \\ 
{-\Theta _{0}-\Theta _{3}-\Upsilon _{0}-\Upsilon _{3}}{=\vartheta _{2,2}}%
\mbox{;} \\ 
{-\Theta _{0}-\Theta _{3}+\Upsilon _{0}+\Upsilon _{3}}{=\vartheta _{3,3}}%
\mbox{;} \\ 
{-\Theta _{0}+\Theta _{3}+\Upsilon _{0}-\Upsilon _{3}}{=\vartheta _{4,4}}%
\end{array}%
\right\vert \mbox{,} 
\]

and $\Theta _1$, $\Upsilon _1$, $\Theta _2$, $\Upsilon _2$, ${M_0}$, ${M_4}$%
, ${M_{\zeta ,0}}$, ${M_{\zeta ,4}}$, ${M_{\eta ,0}}$, ${M_{\eta ,4}}$, ${%
M_{\theta ,0}}$, ${M_{\theta ,4}} $ be solutions of the following sets of
equations: 
\[
\left\{ 
\begin{array}{c}
{\ \Theta _1+\Upsilon _1}{}{=\vartheta _{1,2}}\mbox{;} \\ 
{-\Theta _1+\Upsilon _1}{}{=\vartheta _{3,4}}\mbox{;}%
\end{array}
\right| 
\]
\[
\left\{ 
\begin{array}{c}
{-\Theta _2-\Upsilon _2}{}{=\varpi _{1,2}}\mbox{;} \\ 
{\Theta _2-\Upsilon _2}{}{=\varpi _{3,4}}\mbox{;}%
\end{array}
\right| 
\]
\[
\left\{ 
\begin{array}{c}
{M_0+M_{\theta ,0}}{}{=\vartheta _{1,3}}\mbox{;} \\ 
{M_0-M_{\theta ,0}}{}{=\vartheta _{2,4}}\mbox{;}%
\end{array}
\right| 
\]
\[
\left\{ 
\begin{array}{c}
{M_4+M_{\theta ,4}}{}{=\varpi _{1,3}}\mbox{;} \\ 
{M_4-M_{\theta ,4}}{}{=\varpi _{2,4}}\mbox{;}%
\end{array}
\right| 
\]
\[
\left\{ 
\begin{array}{c}
{M_{\zeta ,0}-M_{\eta ,4}}{}{=\vartheta _{1,4}}\mbox{;} \\ 
{M_{\zeta ,0}+M_{\eta ,4}}{}{=\vartheta _{2,3}}\mbox{;}%
\end{array}
\right| 
\]
\[
\left\{ 
\begin{array}{c}
{M_{\zeta ,4}-M_{\eta ,0}}{}{=\varpi _{1,4}}\mbox{;} \\ 
{M_{\zeta ,4}+M_{\eta ,0}}{}{=\varpi _{2,3}}%
\end{array}
\right|\mbox{.} 
\]

Thus the columns of $\widehat{Q}$ are the following:

the first and the second columns:

\[
\mbox{ } 
\begin{array}{cc}
-\mathrm{i}\Theta _0+\mathrm{i}\Theta _3-\mathrm{i}\Upsilon _0+\mathrm{i}%
\Upsilon _3 & \mathrm{i}\Theta _1+\mathrm{i}\Upsilon _1+\Theta _2+\Upsilon _2
\\ 
\mathrm{i}\Theta _1+\mathrm{i}\Upsilon _1-\Theta _2-\Upsilon _2 & -\mathrm{i}%
\Theta _0-\mathrm{i}\Theta _3-\mathrm{i}\Upsilon _0-\mathrm{i}\Upsilon _3 \\ 
\mathrm{i}M_0+\mathrm{i}M_{\theta ,0}+M_4+M_{\theta ,4} & \mathrm{i}M_{\zeta
,0}+\mathrm{i}M_{\eta ,4}+M_{\zeta ,4}+M_{\eta ,0} \\ 
\mathrm{i}M_{\zeta ,0}-\mathrm{i}M_{\eta ,4}+M_{\zeta ,4}-M_{\eta ,0} & 
\mathrm{i}M_0-\mathrm{i}M_{\theta ,0}+M_4-M_{\theta ,4}%
\end{array}
\mbox{,} 
\]

the third and the fourth columns:

\[
\begin{array}{cc}
\mathrm{i}M_0+\mathrm{i}M_{\theta ,0}-M_4-M_{\theta ,4} & \mathrm{i}M_{\zeta
,0}-\mathrm{i}M_{\eta ,4}-M_{\zeta ,4}+M_{\eta ,0} \\ 
\mathrm{i}M_{\zeta ,0}+\mathrm{i}M_{\eta ,4}-M_{\zeta ,4}-M_{\eta ,0} & 
\mathrm{i}M_0-\mathrm{i}M_{\theta ,0}-M_4+M_{\theta ,4} \\ 
-\mathrm{i}\Theta _0-\mathrm{i}\Theta _3+\mathrm{i}\Upsilon _0+\mathrm{i}%
\Upsilon _3 & -\mathrm{i}\Theta _1+\mathrm{i}\Upsilon _1-\Theta _2+\Upsilon
_2 \\ 
-\mathrm{i}\Theta _1+\mathrm{i}\Upsilon _1+\Theta _2-\Upsilon _2 & -\mathrm{i%
}\Theta _0+\mathrm{i}\Theta _3+\mathrm{i}\Upsilon _0-\mathrm{i}\Upsilon _3%
\end{array}
\mbox{.} 
\]

Hence,

\[
\begin{array}{c}
\widehat{Q}= \\ 
=\mathrm{i}\Theta _0\beta ^{\left[ 0\right] }+\mathrm{i}\Upsilon _0\beta ^{%
\left[ 0\right] }\gamma ^{\left[ 5\right] }+ \\ 
+\mathrm{i}\Theta _1\beta ^{\left[ 1\right] }+\mathrm{i}\Upsilon _1\beta ^{%
\left[ 1\right] }\gamma ^{\left[ 5\right] }+ \\ 
+\mathrm{i}\Theta _2\beta ^{\left[ 2\right] }+\mathrm{i}\Upsilon _2\beta ^{%
\left[ 2\right] }\gamma ^{\left[ 5\right] }+ \\ 
+\mathrm{i}\Theta _3\beta ^{\left[ 3\right] }+\mathrm{i}\Upsilon _3\beta ^{%
\left[ 3\right] }\gamma ^{\left[ 5\right] }+ \\ 
+\mathrm{i}M_0\gamma ^{\left[ 0\right] }+\mathrm{i}M_4\beta ^{\left[ 4\right]
}- \\ 
-\mathrm{i}M_{\zeta ,0}\gamma _\zeta ^{[0]}+\mathrm{i}M_{\zeta ,4}\zeta
^{[4]}- \\ 
-\mathrm{i}M_{\eta ,0}\gamma _\eta ^{[0]}-\mathrm{i}M_{\eta ,4}\eta ^{[4]}+
\\ 
+\mathrm{i}M_{\theta ,0}\gamma _\theta ^{[0]}+\mathrm{i}M_{\theta ,4}\theta
^{[4]}\mbox{.}%
\end{array}
\]

Therefore, from (\ref{ham1}):

\begin{equation}
\frac 1{\mathrm{c}}\partial _t\varphi -\left( \mathrm{i}\Theta _0\beta ^{%
\left[ 0\right] }+\mathrm{i}\Upsilon _0\beta ^{\left[ 0\right] }\gamma ^{%
\left[ 5\right] }\right) \varphi =\left( 
\begin{array}{c}
\sum\limits_{\nu =1}^3\beta ^{\left[ \nu \right] }\left( \partial _\nu +%
\mathrm{i}\Theta _\nu +\mathrm{i}\Upsilon _\nu \gamma ^{\left[ 5\right]
}\right) + \\ 
+\mathrm{i}M_0\gamma ^{\left[ 0\right] }+\mathrm{i}M_4\beta ^{\left[ 4\right]
}- \\ 
-\mathrm{i}M_{\zeta ,0}\gamma _\zeta ^{[0]}+\mathrm{i}M_{\zeta ,4}\zeta
^{[4]}- \\ 
-\mathrm{i}M_{\eta ,0}\gamma _\eta ^{[0]}-\mathrm{i}M_{\eta ,4}\eta ^{[4]}+
\\ 
+\mathrm{i}M_{\theta ,0}\gamma _\theta ^{[0]}+\mathrm{i}M_{\theta ,4}\theta
^{[4]}%
\end{array}
\right) \varphi \mbox{.}  \label{ham0}
\end{equation}

with 
\begin{equation}
\gamma ^{\left[ 5\right] }:=\left[ 
\begin{array}{cc}
1_2 & 0_2 \\ 
0_2 & -1_2%
\end{array}
\right] \mbox{.}  \label{g5}
\end{equation}

Because 
\[
\zeta ^{[k]}+\eta ^{[k]}+\theta ^{[k]}=-\beta ^{\left[ k\right] } 
\]
with $k\in \left\{ 1,2,3\right\} $ then from (\ref{ham0}): 
\begin{eqnarray*}
&&\rule{-1cm}{0pt}\left( 
\begin{array}{c}
-\left( \partial _0+\mathrm{i}\Theta _0+\mathrm{i}\Upsilon _0\gamma ^{\left[
5\right] }\right) +\sum\limits_{k=1}^3\beta ^{\left[ k\right] }\left(
\partial _k+\mathrm{i}\Theta _k+\mathrm{i}\Upsilon _k\gamma ^{\left[ 5\right]
}\right) \\ 
+2\left( \mathrm{i}M_0\gamma ^{\left[ 0\right] }+\mathrm{i}M_4\beta ^{\left[
4\right] }\right)%
\end{array}
\right) \varphi + \\[4pt]
&&\rule{-1cm}{0pt}+\left( 
\begin{array}{c}
-\left( \partial _0+\mathrm{i}\Theta _0+\mathrm{i}\Upsilon _0\gamma ^{\left[
5\right] }\right) -\sum\limits_{k=1}^3\zeta ^{[k]}\left( \partial _k+\mathrm{%
i}\Theta _k+\mathrm{i}\Upsilon _k\gamma ^{\left[ 5\right] }\right) \\ 
+2\left( -\mathrm{i}M_{\zeta ,0}\gamma _\zeta ^{[0]}+\mathrm{i}M_{\zeta
,4}\zeta ^{[4]}\right)%
\end{array}
\right) \varphi + \\[4pt]
&&\rule{-1cm}{0pt}+\left( 
\begin{array}{c}
\left( \partial _0+\mathrm{i}\Theta _0+\mathrm{i}\Upsilon _0\gamma ^{\left[ 5%
\right] }\right) -\sum\limits_{k=1}^3\eta ^{[k]}\left( \partial _k+\mathrm{i}%
\Theta _k+\mathrm{i}\Upsilon _k\gamma ^{\left[ 5\right] }\right) \\ 
+2\left( -\mathrm{i}M_{\eta ,0}\gamma _\eta ^{[0]}-\mathrm{i}M_{\eta ,4}\eta
^{[4]}\right)%
\end{array}
\right) \varphi + \\[4pt]
&&\rule{-1cm}{0pt}+\left( 
\begin{array}{c}
-\left( \partial _0+\mathrm{i}\Theta _0+\mathrm{i}\Upsilon _0\gamma ^{\left[
5\right] }\right) -\sum\limits_{k=1}^3\theta ^{[k]}\left( \partial _k+%
\mathrm{i}\Theta _k+\mathrm{i}\Upsilon _k\gamma ^{\left[ 5\right] }\right)
\\ 
+2\left( \mathrm{i}M_{\theta ,0}\gamma _\theta ^{[0]}+\mathrm{i}M_{\theta
,4}\theta ^{[4]}\right)%
\end{array}
\right) \varphi =0\mbox{.}
\end{eqnarray*}

\section{Quarks and Gluons}

The following part of (\ref{ham0}):
\begin{equation}
\left( 
\begin{array}{c}
\sum\limits_{k=0}^3\beta ^{\left[ k\right] }\left( -\mathrm{i}\partial
_k+\Theta _k+\Upsilon _k\gamma ^{\left[ 5\right] }\right) - \\[2pt]
-M_{\zeta ,0}\gamma _\zeta ^{[0]}+M_{\zeta ,4}\zeta ^{[4]}\,+ \\[2pt]
-M_{\eta ,0}\gamma _\eta ^{[0]}-M_{\eta ,4}\eta ^{[4]}\,+ \\[2pt]
+M_{\theta ,0}\gamma _\theta ^{[0]}+M_{\theta ,4}\theta ^{[4]}
\end{array}
\right) \varphi =0\mbox{.}  \label{clrH}
\end{equation}
is called {\it the chromatic equation of moving}.

Here (\ref{redM}), (\ref{greenM}), (\ref{blueM}):
\[
\gamma _\zeta ^{[0]}=-\left[ 
\begin{array}{cccc}
0 & 0 & 0 & 1 \\ 
0 & 0 & 1 & 0 \\ 
0 & 1 & 0 & 0 \\ 
1 & 0 & 0 & 0
\end{array}
\right] , \ \ \zeta ^{[4]}=\left[ 
\begin{array}{cccc}
0 & 0 & 0 & \mathrm{i} \\ 
0 & 0 & \mathrm{i} & 0 \\ 
0 & -\mathrm{i} & 0 & 0 \\ 
-\mathrm{i} & 0 & 0 & 0
\end{array}
\right] 
\]
are mass elements of red pentad;
\[
\gamma _\eta ^{[0]}=\left[ 
\begin{array}{cccc}
0 & 0 & 0 & \mathrm{i} \\ 
0 & 0 & -\mathrm{i} & 0 \\ 
0 & \mathrm{i} & 0 & 0 \\ 
-\mathrm{i} & 0 & 0 & 0
\end{array}
\right] , \ \ \eta ^{[4]}=\left[ 
\begin{array}{cccc}
0 & 0 & 0 & 1 \\ 
0 & 0 & -1 & 0 \\ 
0 & -1 & 0 & 0 \\ 
1 & 0 & 0 & 0
\end{array}
\right] 
\]
are mass elements of green pentad;
\[
\gamma _{\theta }^{[0]}=\left[ 
\begin{array}{cccc}
0 & 0 & -1 & 0 \\ 
0 & 0 & 0 & 1 \\ 
-1 & 0 & 0 & 0 \\ 
0 & 1 & 0 & 0%
\end{array}%
\right] ,\ \ \theta ^{\lbrack 4]}=\left[ 
\begin{array}{cccc}
0 & 0 & \mathrm{i} & 0 \\ 
0 & 0 & 0 & -\mathrm{i} \\ 
-\mathrm{i} & 0 & 0 & 0 \\ 
0 & \mathrm{i} & 0 & 0%
\end{array}%
\right] 
\]%
are mass elements of blue pentad.

I call:
\begin{itemize}
\item
$M_{\zeta ,0}$, $M_{\zeta ,4}$ {\it red lower and upper mass members};
\item
$M_{\eta ,0}$, $M_{\eta ,4}$ {\it green lower and upper mass members};
\item
$M_{\theta ,0}$, $M_{\theta ,4}$ {\it blue lower and upper mass members}.
\end{itemize}

The mass members of this equation form the following matrix sum:
\vspace*{-2pt}
\[
\widehat{M}:= 
\left(\begin{array}{c}
-\,M_{\zeta ,0}\gamma _\zeta ^{[0]}+M_{\zeta ,4}\zeta ^{[4]}\,- \\[3pt]
-\,M_{\eta ,0}\gamma _\eta ^{[0]}-M_{\eta ,4}\eta ^{[4]}\,+ \\[3pt]
+\,M_{\theta ,0}\gamma _\theta ^{[0]}+M_{\theta ,4}\theta ^{[4]}
\end{array}\right) =
\]
\begin{eqnarray*}
&&\rule{-1cm}{0pt}=\ \left[ 
\begin{array}{cccc}
0 & 0 & -M_{\theta ,0} & M_{\zeta ,\eta ,0} \\[3pt]
0 & 0 & M_{\zeta ,\eta ,0}^{*} & M_{\theta ,0} \\[3pt]
-M_{\theta ,0} & M_{\zeta ,\eta ,0} & 0 & 0 \\[3pt]
M_{\zeta ,\eta ,0}^{*} & M_{\theta ,0} & 0 & 0
\end{array}
\right] +\mathrm{i}\left[ 
\begin{array}{cccc}
0 & 0 & M_{\theta ,4} & M_{\zeta ,\eta ,4}^{*} \\[3pt]
0 & 0 & M_{\zeta ,\eta ,4} & -M_{\theta ,4} \\[3pt]
-M_{\theta ,4} & -M_{\zeta ,\eta ,4}^{*} & 0 & 0 \\[3pt]
-M_{\zeta ,\eta ,4} & M_{\theta ,4} & 0 & 0
\end{array}
\right] 
\end{eqnarray*}

with $M_{\zeta ,\eta ,0}:=M_{\zeta ,0}-\mathrm{i}M_{\eta ,0}$ and $M_{\zeta
,\eta ,4}:=M_{\zeta ,4}-\mathrm{i}M_{\eta ,4}$.

Elements of these matrices can be turned by formula of shape \cite{Z}:
\[
\begin{array}{c}
\left[ 
\begin{array}{cc}
\cos \frac \theta 2 & \mathrm{i}\sin \frac \theta 2 \\[4pt]
\mathrm{i}\sin \frac \theta 2 & \cos \frac \theta 2
\end{array}
\right] \!\left[ 
\begin{array}{cc}
Z & X-\mathrm{i}Y \\[4pt]
X+\mathrm{i}Y & -Z
\end{array}
\right] \left[ 
\begin{array}{cc}
\cos \frac \theta 2 & -\mathrm{i}\sin \frac \theta 2 \\[4pt]
-\mathrm{i}\sin \frac \theta 2 & \cos \frac \theta 2
\end{array}
\right] = \\ 
=\left[ 
\begin{array}{cc}
Z\cos \theta -Y\sin \theta  & X-\mathrm{i}\left( 
\begin{array}{c}
Y\cos \theta  \\ 
+Z\sin \theta 
\end{array}
\right)  \\[6pt]
X+\mathrm{i}\left( 
\begin{array}{c}
Y\cos \theta  \\ 
+Z\sin \theta 
\end{array}
\right)  & -Z\cos \theta +Y\sin \theta 
\end{array}
\right] 
\end{array}\mbox{.}
\]

Hence, if:
\[
\widehat{M}^{\prime }:= 
\left(\!\!\begin{array}{c}
-M_{\zeta ,0}^{\prime }\gamma _\zeta ^{[0]}+M_{\zeta ,4}^{\prime }\zeta
^{[4]}- \\[3pt]
-M_{\eta ,0}^{\prime }\gamma _\eta ^{[0]}-M_{\eta ,4}^{\prime }\eta ^{[4]}+
\\[3pt]
+M_{\theta ,0}^{\prime }\gamma _\theta ^{[0]}+M_{\theta ,4}^{\prime }\theta
^{[4]}
\end{array}\!\!\right)\!
:=U_{2,3}^{\dagger }\left( \alpha \right) \widehat{M}%
U_{2,3}\left( \alpha \right) 
\]
then
\vspace*{-4pt}
\begin{eqnarray*}
&&\rule{-1cm}{0pt}
M_{\zeta ,0}^{\prime }=M_{\zeta ,0}\,,
\\
&&\rule{-1cm}{0pt}
M_{\eta ,0}^{\prime }=M_{\eta ,0}\cos 2\alpha +M_{\theta ,0}\sin 2\alpha \,,
\\
&&\rule{-1cm}{0pt}
M_{\theta ,0}^{\prime }=M_{\theta ,0}\cos 2\alpha -M_{\eta ,0}\sin 2\alpha \,,
\\
&&\rule{-1cm}{0pt}
M_{\zeta ,4}^{\prime }=M_{\zeta ,4}\,,
\\
&&\rule{-1cm}{0pt}
M_{\eta ,4}^{\prime }=M_{\eta ,4}\cos 2\alpha +M_{\theta ,4}\sin 2\alpha \,,
\\
&&\rule{-1cm}{0pt}
M_{\theta ,4}^{\prime }=M_{\theta ,4}\cos 2\alpha -M_{\eta ,4}\sin 2\alpha \,.
\end{eqnarray*}

Therefore, matrix $U_{2,3}\left( \alpha \right) $ makes an oscillation
between green and blue chromatics.

Let us consider equation (\ref{ham0}) under transformation $U_{2,3}\left(
\alpha \right) $ where $\alpha $ is an arbitrary real function of time-space
variables ($\alpha =\alpha \left( t,x_1,x_2,x_3\right) $):
\begin{eqnarray*}
&&\rule{-.8cm}{0pt}U_{2,3}^{\dagger }\left( \alpha \right) \left( \frac 1{%
\mathrm{c}}\,\partial _t+\mathrm{i}\Theta _0+\mathrm{i}\Upsilon _0\gamma
^{\left[ 5\right] }\right) U_{2,3}\left( \alpha \right) \varphi = \\
&&\rule{-.8cm}{0pt}=U_{2,3}^{\dagger }\left( \alpha \right) \left( \!\!
\begin{array}{c}
\sum\limits_{\nu =1}^3\beta ^{\left[ \nu \right] }\left( \partial _\nu +%
\mathrm{i}\Theta _\nu +\mathrm{i}\Upsilon _\nu \gamma ^{\left[ 5\right]
}\right) + \\[3pt]
+\,\mathrm{i}M_0\gamma ^{\left[ 0\right] }+\mathrm{i}M_4\beta ^{\left[
4\right] }+\widehat{M}
\end{array}
\!\!\right) U_{2,3}\left( \alpha \right) \varphi \,\mbox{.}
\end{eqnarray*}

Because
\vspace*{-5pt}
\begin{eqnarray*}
&
U_{2,3}^{\dagger }\left( \alpha \right) U_{2,3}\left( \alpha \right) =1_4\,,
\\
&
U_{2,3}^{\dagger }\left( \alpha \right) \gamma ^{\left[ 5\right]
}U_{2,3}\left( \alpha \right) =\gamma ^{\left[ 5\right] }\,,
\\[6pt]
&
U_{2,3}^{\dagger }\left( \alpha \right) \gamma ^{\left[ 0\right]
}U_{2,3}\left( \alpha \right) =\gamma ^{\left[ 0\right] }\,,
\\
&
U_{2,3}^{\dagger }\left( \alpha \right) \beta ^{\left[ 4\right]
}U_{2,3}\left( \alpha \right) =\beta ^{\left[ 4\right] }\,,
\\[6pt]
&
U_{2,3}^{\dagger }\left( \alpha \right) \beta ^{\left[ 1\right] }=\beta
^{\left[ 1\right] }U_{2,3}^{\dagger }\left( \alpha \right)\,,
\\[6pt]
&
U_{2,3}^{\dagger }\left( \alpha \right) \beta ^{\left[ 2\right] }=\left(
\beta ^{\left[ 2\right] }\cos 2\alpha +\beta ^{\left[ 3\right] }\sin 2\alpha
\right) U_{2,3}^{\dagger }\left( \alpha \right)\,,
\\
&
U_{2,3}^{\dagger }\left( \alpha \right) \beta ^{\left[ 3\right] }=\left(
\beta ^{\left[ 3\right] }\cos 2\alpha -\beta ^{\left[ 2\right] }\sin 2\alpha
\right) U_{2,3}^{\dagger }\left( \alpha \right)\,,
\end{eqnarray*}
then
 \begin{eqnarray}
&&\rule{-1.3cm}{0pt}\left( \frac 1{\mathrm{c}}\,\partial _t+U_{2,3}^{\dagger
}\left( \alpha \right) \frac 1{\mathrm{c}}\,\partial _tU_{2,3}\left( \alpha
\right) +\mathrm{i}\Theta _0+\mathrm{i}\Upsilon _0\gamma ^{\left[ 5\right]
}\right) \varphi =  \nonumber \\[3pt]
&&\rule{-1.3cm}{0pt}=\left( 
\begin{array}{c}
\beta ^{\left[ 1\right] }\left( \!\!\!\partial _1+U_{2,3}^{\dagger }\left(
\alpha \right) \partial _1U_{2,3}\left( \alpha \right) +\,\mathrm{i}\Theta
_1+\mathrm{i}\Upsilon _1\gamma ^{\left[ 5\right] }\!\!\!\right)+ \, \\
+\beta ^{\left[ 2\right] }\!\!\left( 
\begin{array}{c}
\left( \cos 2\alpha \cdot \partial _2-\sin 2\alpha \cdot \partial _3\right) 
\\[3pt]
+\,U_{2,3}^{\dagger }\left( \alpha \right) \left( \!\!\!\cos 2\alpha \cdot
\partial _2-\,\sin 2\alpha \cdot \partial _3\!\!\!\right) U_{2,3}\left(
\alpha \right)  \\[3pt]
+\,\mathrm{i}\left( \Theta _2\cos 2\alpha -\Theta _3\sin 2\alpha \right)  \\%
[3pt]
+\,\mathrm{i}\left( \Upsilon _2\gamma ^{\left[ 5\right] }\cos 2\alpha
-\Upsilon _3\gamma ^{\left[ 5\right] }\sin 2\alpha \right) 
\end{array}
\!\!\!\right)  \\ 
+\,\beta ^{\left[ 3\right] }\!\left( \!\!\!
\begin{array}{c}
\left( \cos 2\alpha \cdot \partial _3+\sin 2\alpha \cdot \partial _2\right) 
\\[3pt]
+\,U_{2,3}^{\dagger }\left( \alpha \right) \left( \!\!\!\cos 2\alpha \cdot
\partial _3+\,\sin 2\alpha \cdot \partial _2\!\!\!\right) U_{2,3}\left(
\alpha \right)  \\[3pt]
+\,\mathrm{i}\left( \Theta _2\sin 2\alpha +\Theta _3\cos 2\alpha \right)  \\%
[3pt]
+\,\mathrm{i}\left( \Upsilon _3\gamma ^{\left[ 5\right] }\cos 2\alpha
+\Upsilon _2\gamma ^{\left[ 5\right] }\sin 2\alpha \right) 
\end{array}
\!\!\!\right)  \\[36pt]
+\,\mathrm{i}M_0\gamma ^{\left[ 0\right] }+\mathrm{i}M_4\beta ^{\left[
4\right] }+\widehat{M}^{\prime }
\end{array}
\!\!\!\right) \varphi \,\mbox{.}  \label{ham5}
\end{eqnarray}

Let $x_2^{\prime }$ and $x_3^{\prime }$ be elements of other coordinate 
system such that \ref{u23}:
\vspace*{-3pt}
\begin{eqnarray}
\partial _2^{\prime } &:&=\left( \cos 2\alpha \cdot \partial _2-\sin 2\alpha
\cdot \partial _3\right) \mbox{,}   \\
\partial _3^{\prime } &:&=\left( \cos 2\alpha \cdot \partial _3+\sin 2\alpha
\cdot \partial _2\right) \mbox{.}  \nonumber
\end{eqnarray}

Therefore, from (\ref{ham5}):
\begin{eqnarray*}
&&\rule{-1.3cm}{0pt}\left( \frac 1{\mathrm{c}}\,\partial _t+U_{2,3}^{\dagger
}\left( \alpha \right) \frac 1{\mathrm{c}}\,\partial _tU_{2,3}\left( \alpha
\right) +\mathrm{i}\Theta _0+\mathrm{i}\Upsilon _0\gamma ^{\left[ 5\right]
}\right) \varphi = \\[3pt]
&&\rule{-1.3cm}{0pt}=\left( 
\begin{array}{c}
\beta ^{\left[ 1\right] }\left( \partial _1+U_{2,3}^{\dagger }\left( \alpha
\right) \partial _1U_{2,3}\left( \alpha \right) +\mathrm{i}\Theta _1+\mathrm{%
i}\Upsilon _1\gamma ^{\left[ 5\right] }\right)  \\[6pt]
+\,\beta ^{\left[ 2\right] }\left( \partial _2^{\prime }+U_{2,3}^{\dagger
}\left( \alpha \right) \partial _2^{\prime }U_{2,3}\left( \alpha \right) +%
\mathrm{i}\Theta _2^{\prime }+\mathrm{i}\Upsilon _2^{\prime }\gamma ^{\left[
5\right] }\right)  \\[6pt]
+\,\beta ^{\left[ 3\right] }\left( \partial _3^{\prime }+U_{2,3}^{\dagger
}\left( \alpha \right) \partial _3^{\prime }U_{2,3}\left( \alpha \right) +%
\mathrm{i}\Theta _3^{\prime }+\mathrm{i}\Upsilon _3^{\prime }\gamma ^{\left[
5\right] }\right)  \\[6pt]
+\,\mathrm{i}M_0\gamma ^{\left[ 0\right] }+\mathrm{i}M_4\beta ^{\left[
4\right] }+\widehat{M}^{\prime }
\end{array}
\right) \varphi \,.
\end{eqnarray*}
with
\[
\begin{array}{c}
\Theta _2^{\prime }:=\Theta _2\cos 2\alpha -\Theta _3\sin
2\alpha \,\mbox{,} \\[4pt]
\Theta _3^{\prime }:=\Theta _2\sin 2\alpha +\Theta _3\cos
2\alpha \,\mbox{,} \\[4pt] 
\Upsilon _2^{\prime }:=\Upsilon _2\cos 2\alpha -\Upsilon
_3\sin 2\alpha \,\mbox{,} \\[4pt]
\Upsilon _3^{\prime }:=\Upsilon _3\cos 2\alpha +\Upsilon
_2\sin 2\alpha \,\mbox{.}
\end{array}
\]

Therefore, the oscillation between blue and green chromatics curves the
space in the $x_2$, $x_3$ directions.

Similarly, matrix $U_{1,3}$ with an arbitrary real function $\vartheta \left( t,x_1,x_2,x_3\right) $
describes the oscillation between blue and red chromatics which curves
the space in the $x_1$, $x_3$ directions. And matrix $U_{1,2}$ with an arbitrary real function $\varsigma \left( t,x_1,x_2,x_3\right) $
describes the oscillation between green and red chromatics which curves
the space in the $x_1$, $x_2$ directions.

Now, let
\[
\widehat{M}^{\prime \prime }:= \!
\left(\!\!\begin{array}{c}
-M_{\zeta ,0}^{\prime \prime }\gamma _\zeta ^{[0]}+M_{\zeta ,4}^{\prime
\prime }\zeta ^{[4]}- \\ 
-M_{\eta ,0}^{\prime \prime }\gamma _\eta ^{[0]}-M_{\eta ,4}^{\prime \prime
}\eta ^{[4]}+ \\ 
+M_{\theta ,0}^{\prime \prime }\gamma _\theta ^{[0]}+M_{\theta ,4}^{\prime
\prime }\theta ^{[4]}
\end{array}\!\!\right)\!
:=U_{0,1}^{\dagger }\left( \sigma \right) \widehat{M}%
U_{0,1}\left( \sigma \right) 
\]
then:
\vspace*{-6pt}
\begin{eqnarray*}
&&\rule{-.5cm}{0pt}
M_{\zeta ,0}^{\prime \prime }=M_{\zeta ,0}\,,
\\[2pt]
&&\rule{-.5cm}{0pt}
M_{\eta ,0}^{\prime \prime }=\left( M_{\eta ,0}\cosh 2\sigma -M_{\theta
,4}\sinh 2\sigma \right)\,,
\\[2pt]
&&\rule{-.5cm}{0pt}
M_{\theta ,0}^{\prime \prime }=M_{\theta ,0}\cosh 2\sigma +M_{\eta
,4}\sinh 2\sigma\,,
\\[2pt]
&&\rule{-.5cm}{0pt}
M_{\zeta ,4}^{\prime \prime }=M_{\zeta ,4}\,,
\\[2pt]
&&\rule{-.5cm}{0pt}
M_{\eta ,4}^{\prime \prime }=M_{\eta ,4}\cosh 2\sigma +M_{\theta ,0}\sinh
2\sigma\,,
\\[2pt]
&&\rule{-.5cm}{0pt}
M_{\theta ,4}^{\prime \prime }=M_{\theta ,4}\cosh 2\sigma -M_{\eta
,0}\sinh 2\sigma\,.
\end{eqnarray*}

Therefore, matrix $U_{0,1}\left( \sigma \right) $ makes an oscillation
between green and blue chromatics with an oscillation between upper and lower
mass members.

Let us consider equation (\ref{ham0}) under transformation $U_{0,1}\left(
\sigma \right) $ where $\sigma $ is an arbitrary real function of
time-space variables ($\sigma =\sigma \left( t,x_1,x_2,x_3\right) $):
\begin{eqnarray*}
&&\rule{-0.5cm}{0pt}U_{0,1}^{\dagger }\left( \sigma \right) \left( \frac 1{%
\mathrm{c}}\,\partial _t+\mathrm{i}\Theta _0+\mathrm{i}\Upsilon _0\gamma
^{\left[ 5\right] }\right) U_{0,1}\left( \sigma \right) \varphi = \\
&&\rule{-0.5cm}{0pt}=U_{0,1}^{\dagger }\left( \sigma \right) \left( \!\!\!
\begin{array}{c}
\sum\limits_{\nu =1}^3\beta ^{\left[ \nu \right] }\left( \partial _\nu +%
\mathrm{i}\Theta _\nu +\mathrm{i}\Upsilon _\nu \gamma ^{\left[ 5\right]
}\right) + \\[3pt]
+\,\mathrm{i}M_0\gamma ^{\left[ 0\right] }+\mathrm{i}M_4\beta ^{\left[
4\right] }+\widehat{M}
\end{array}
\!\!\!\right) U_{0,1}\left( \sigma \right) \varphi \,.
\end{eqnarray*}

Since:
\begin{eqnarray*}
&&U_{0,1}^{\dagger }\left( \sigma \right) U_{0,1}\left( \sigma \right)
=\left( \cosh 2\sigma -\beta ^{\left[ 1\right] }\sinh 2\sigma \right) %
\mbox{,} \\
&&U_{0,1}^{\dagger }\left( \sigma \right) =\left( \cosh 2\sigma +\beta
^{\left[ 1\right] }\sinh 2\sigma \right) U_{0,1}^{-1}\left( \sigma
\right) \mbox{,} \\
&&U_{0,1}^{\dagger }\left( \sigma \right) \beta ^{\left[ 1\right] }
=\left( \beta ^{\left[ 1\right] }\cosh 2\sigma -\sinh 2\sigma \right)
U_{0,1}^{-1}\left( \sigma \right) \mbox{,} \\
&&U_{0,1}^{\dagger }\left( \sigma \right) \beta ^{\left[ 2\right] } =\beta
^{\left[ 2\right] }U_{0,1}^{-1}\left( \sigma \right) \mbox{,} \\[3pt]
&&U_{0,1}^{\dagger }\left( \sigma \right) \beta ^{\left[ 3\right] } =\beta
^{\left[ 3\right] }U_{0,1}^{-1}\left( \sigma \right) \mbox{,} \\[3pt]
&&U_{0,1}^{\dagger }\left( \sigma \right) \gamma ^{\left[ 0\right]
}U_{0,1}\left( \sigma \right) =\gamma ^{\left[ 0\right] }\mbox{,} \\[3pt]
&&U_{0,1}^{\dagger }\left( \sigma \right) \beta ^{\left[ 4\right]
}U_{0,1}\left( \sigma \right) =\beta ^{\left[ 4\right] } \mbox{,}\\[3pt]
&&U_{0,1}^{-1}\left( \sigma \right) U_{0,1}\left( \sigma \right) =1_4\,
\mbox{,} \\[3pt]
&&U_{0,1}^{-1}\left( \sigma \right) \gamma ^{\left[ 5\right] }U_{0,1}\left(
\sigma \right) =\gamma ^{\left[ 5\right] }\,\mbox{,} \\[3pt]
&&U_{0,1}^{\dagger }\left( \sigma \right) \gamma ^{\left[ 5\right]
}U_{0,1}\left( \sigma \right) =\gamma ^{\left[ 5\right] }\left( \cosh
2\sigma -\beta ^{\left[ 1\right] }\sinh 2\sigma \right), 
\end{eqnarray*}
then
\begin{equation}
\left( \!\!\!\!
\begin{array}{c}
U_{0,1}^{-1}\left( \sigma \right) \left( \!\!\!\cosh 2\sigma \cdot \frac 1{%
\mathrm{c}}\,\partial _t+\sinh 2\sigma \cdot \partial _1\!\!\!\right)
U_{0,1}\left( \sigma \right)  \\[13pt]
+\,\left( \cosh 2\sigma \cdot \frac 1{\mathrm{c}}\,\partial _t+\sinh 2\sigma
\cdot \partial _1\right)  \\[3pt]
+\,\mathrm{i}\left( \Theta _0\cosh 2\sigma +\Theta _1\sinh 2\sigma \right) 
\\[3pt]
+\,\mathrm{i}\left( \Upsilon _0\cosh 2\sigma +\sinh 2\sigma \cdot \Upsilon
_1\right) \gamma ^{\left[ 5\right] }- \\[4pt]
-\,\beta ^{\left[ 1\right] }\!\left( \!\!\!
\begin{array}{c}
U_{0,1}^{-1}\left( \sigma \right) \left( \!\!\!\cosh 2\sigma \cdot \partial
_1+\sinh 2\sigma \cdot \frac 1{\mathrm{c}}\,\partial _t\!\!\!\right)
U_{0,1}\left( \sigma \right)  \\[13pt]
+\,\left( \cosh 2\sigma \cdot \partial _1+\sinh 2\sigma \cdot \frac 1{%
\mathrm{c}}\partial _t\right)  \\[3pt]
+\,\mathrm{i}\left( \Theta _1\cosh 2\sigma +\Theta _0\sinh 2\sigma \right) 
\\[3pt]
+\,\mathrm{i}\left( \Upsilon _1\cosh 2\sigma +\Upsilon _0\sinh 2\sigma
\right) \gamma ^{\left[ 5\right] }
\end{array}
\!\!\!\right)  \\[38pt]
-\,\beta ^{\left[ 2\right] }\left( \!\!\!\partial _2+U_{0,1}^{-1}\left(
\sigma \right) \left( \partial _2U_{0,1}\left( \sigma \right) \right) +\,%
\mathrm{i}\Theta _2+\mathrm{i}\Upsilon _2\gamma ^{\left[ 5\right]
}\!\!\!\right)  \\[13pt]
-\,\beta ^{\left[ 3\right] }\left( \!\!\!\partial _3+U_{0,1}^{-1}\left(
\sigma \right) \left( \partial _3U_{0,1}\left( \sigma \right) \right) +\,%
\mathrm{i}\Theta _3+\mathrm{i}\Upsilon _3\gamma ^{\left[ 5\right]
}\!\!\!\right)  \\[13pt]
-\,\mathrm{i}M_0\gamma ^{\left[ 0\right] }-\mathrm{i}M_4\beta ^{\left[
4\right] }-\widehat{M}^{\prime \prime }
\end{array}
\!\!\!\!\right) \varphi =0\,.  \label{ham6}
\end{equation}

Let $t^{\prime }$ and $x_1^{\prime }$ be elements of other coordinate system such that:
\vspace*{2pt}
\begin{equation}
\left.\begin{array}{ll}
\displaystyle \frac{\partial x_1}{\partial x_1^{\prime }}=\cosh 2\sigma 
 \\[10pt]
\displaystyle \frac{\partial t}{\partial x_1^{\prime }}=\frac 1{\mathrm{c}}\sinh
2\sigma  \\[10pt]
\displaystyle \frac{\partial x_1}{\partial t^{\prime }}=\mathrm{c}\sinh 2\sigma %
\\[10pt]
\displaystyle \frac{\partial t}{\partial t^{\prime }}=\cosh 2\sigma
\\[10pt]
\displaystyle \frac{\partial x_2}{\partial t^{\prime }}=\frac{\partial x_3}{\partial
t^{\prime }}=\frac{\partial x_2}{\partial x_1^{\prime }}=\frac{\partial x_3}{%
\partial x_1^{\prime }}=0
\end{array}
\right\}.
\label{grg}
\end{equation}

Hence:
\begin{eqnarray*}
&&\rule{-0.6cm}{0pt}
\partial _t^{\prime }:=\frac \partial {\partial t^{\prime }}
=\frac \partial {\partial t}\frac{\partial t}{\partial t^{\prime }}+\frac
\partial {\partial x_1}\frac{\partial x_1}{\partial t^{\prime }}+\frac
\partial {\partial x_2}\frac{\partial x_2}{\partial t^{\prime }}+\frac
\partial {\partial x_3}\frac{\partial x_3}{\partial t^{\prime }}= \\
&&\rule{-0.6cm}{0pt}
=\cosh 2\sigma \cdot \frac \partial {\partial t}+\mathrm{c}\sinh
2\sigma \cdot \frac \partial {\partial x_1} =\\
&&\rule{-0.6cm}{0pt}
=\cosh 2\sigma \cdot \partial _t+\mathrm{c}\sinh 2\sigma \cdot
\partial _1\,,
\end{eqnarray*}
that is
\[
\frac 1{\mathrm{c}}\,\partial _t^{\prime }=\frac 1{\mathrm{c}}\,\cosh 2\sigma
\cdot \partial _t+\sinh 2\sigma \cdot \partial _1 
\]
and
\begin{eqnarray*}
&&\rule{-.8cm}{0pt}
\partial _1^{\prime } :=\frac \partial {\partial x_1^{\prime }}= \\
&&\rule{-.8cm}{0pt}
=\frac \partial {\partial t}\frac{\partial t}{\partial x_1^{\prime }}%
+\frac \partial {\partial x_1}\frac{\partial x_1}{\partial x_1^{\prime }}%
+\frac \partial {\partial x_2}\frac{\partial x_2}{\partial x_1^{\prime }}%
+\frac \partial {\partial x_3}\frac{\partial x_3}{\partial x_1^{\prime }} =\\
&&\rule{-.8cm}{0pt}
=\cosh 2\sigma \cdot \frac \partial {\partial x_1}+\sinh 2\sigma \cdot
\frac 1{\mathrm{c}}\,\frac \partial {\partial t}= \\
&&\rule{-.8cm}{0pt}
=\cosh 2\sigma \cdot \partial _1+\sinh 2\sigma \cdot \frac 1{\mathrm{c}%
}\,\partial _t\,.
\end{eqnarray*}

Therefore, from (\ref{ham6}):
\[
\left( 
\begin{array}{c}
\beta ^{\left[ 0\right] }\left( \frac 1{\mathrm{c}}\,\partial _t^{\prime
}+U_{0,1}^{-1}\left( \sigma \right) \frac 1{\mathrm{c}}\,\partial _t^{\prime
}U_{0,1}\left( \sigma \right) +\,\mathrm{i}\Theta _0^{\prime \prime }+%
\mathrm{i}\Upsilon _0^{\prime \prime }\gamma ^{\left[ 5\right] }\right)  \\%
[13pt]
+\,\beta ^{\left[ 1\right] }\left( \partial _1^{\prime }+U_{0,1}^{-1}\left(
\sigma \right) \partial _1^{\prime }U_{0,1}\left( \sigma \right) +\,\mathrm{i%
}\Theta _1^{\prime \prime }+\mathrm{i}\Upsilon _1^{\prime \prime }\gamma
^{\left[ 5\right] }\right)  \\[13pt]
+\,\beta ^{\left[ 2\right] }\left( \partial _2+U_{0,1}^{-1}\left( \sigma
\right) \partial _2U_{0,1}\left( \sigma \right) +\,\mathrm{i}\Theta _2+%
\mathrm{i}\Upsilon _2\gamma ^{\left[ 5\right] }\right)  \\[13pt]
+\,\beta ^{\left[ 3\right] }\left( \partial _3+U_{0,1}^{-1}\left( \sigma
\right) \partial _3U_{0,1}\left( \sigma \right) +\,\mathrm{i}\Theta _3+%
\mathrm{i}\Upsilon _3\gamma ^{\left[ 5\right] }\right)  \\[13pt]
+\,\mathrm{i}M_0\gamma ^{\left[ 0\right] }+\mathrm{i}M_4\beta ^{\left[
4\right] }+\widehat{M}^{\prime \prime }
\end{array}
\right) \varphi =0
\]

with
\[
\begin{array}{ll}
\Theta _0^{\prime \prime }:=\Theta _0\cosh 2\sigma +\Theta
_1\sinh 2\sigma \,, \\[4pt]
\Theta _1^{\prime \prime }:=\Theta _1\cosh 2\sigma +\Theta
_0\sinh 2\sigma \,, \\[4pt]
\Upsilon _0^{\prime \prime }:=\Upsilon _0\cosh 2\sigma
+\sinh 2\sigma \cdot \Upsilon _1\,, \\[4pt]
\Upsilon _1^{\prime \prime }:=\Upsilon _1\cosh 2\sigma
+\Upsilon _0\sinh 2\sigma \,.
\end{array}
\]

Therefore, the oscillation between blue and green chromatics with the
oscillation between upper and lower mass members curves the space in the $t$, $x_1$ directions.

Similarly, matrix $U_{0,2}$ with an arbitrary real function $\phi \left( t,x_1,x_2,x_3\right) $
describes the oscillation between blue and red chromatics with the
oscillation between upper and lower mass members curves the space in
the $t$, $x_2$ directions. And matrix $U_{0,3}$ with an arbitrary real function $\iota \left( t,x_1,x_2,x_3\right) $
describes the oscillation between green and red chromatics with the
oscillation between upper and lower mass members curves the space in
the $t$, $x_3$ directions.

Now let
\[
\widetilde{U}\left( \chi \right) :=\left[ 
\begin{array}{cccc}
e^{\mathrm{i}\chi } & 0 & 0 & 0 \\ 
0 & e^{\mathrm{i}\chi } & 0 & 0 \\ 
0 & 0 & e^{2\mathrm{i}\chi } & 0 \\ 
0 & 0 & 0 & e^{2\mathrm{i}\chi }
\end{array}
\right] 
\]
and
\[
\widehat{M}^{\prime }:= 
\left(\begin{array}{c}
-\,M_{\zeta ,0}^{\prime }\gamma _\zeta ^{[0]}+M_{\zeta ,4}^{\prime }\zeta
^{[4]}\,- \\[3pt]
-M_{\eta ,0}^{\prime }\gamma _\eta ^{[0]}-M_{\eta ,4}^{\prime }\eta ^{[4]}\,+
\\[3pt]
+\,M_{\theta ,0}^{\prime }\gamma _\theta ^{[0]}+M_{\theta ,4}^{\prime }\theta
^{[4]}
\end{array}\right)
:=\widetilde{U}^{\dagger }\left( \chi \right) \widehat{M}%
\widetilde{U}\left( \chi \right) 
\]
then:
\vspace*{-5pt}
\begin{eqnarray*}
M_{\zeta ,0}^{\prime } \!\!&\!\!{=}\!\!&\!\!\left( M_{\zeta ,0}\cos \chi -M_{\zeta ,4}\sin
\chi \right) \mbox{,} \\[3pt]
M_{\zeta ,4}^{\prime } \!\!&\!\!{=}\!\!&\!\!\left( M_{\zeta ,4}\cos \chi +M_{\zeta ,0}\sin
\chi \right) \mbox{,} \\[3pt]
M_{\eta ,4}^{\prime } \!\!&\!\!{=}\!\!&\!\!\left( M_{\eta ,4}\cos \chi -M_{\eta ,0}\sin \chi
\right) \mbox{,} \\[3pt]
M_{\eta ,0}^{\prime } \!\!&\!\!{=}\!\!&\!\!\left( M_{\eta ,0}\cos \chi +M_{\eta ,4}\sin \chi
\right) \mbox{,} \\[3pt]
M_{\theta ,0}^{\prime } \!\!&\!\!{=}\!\!&\!\!\left( M_{\theta ,0}\cos \chi +M_{\theta ,4}\sin
\chi \right) \mbox{,} \\[3pt]
M_{\theta ,4}^{\prime } \!\!&\!\!{=}\!\!&\!\!\left( M_{\theta ,4}\cos \chi -M_{\theta ,0}\sin
\chi \right) \mbox{.}
\end{eqnarray*}

Therefore, matrix $\widetilde{U}\left( \chi \right) $ makes an oscillation
between upper and lower mass members.

Let us consider equation (\ref{clrH}) under transformation $\widetilde{U}%
\left( \chi \right) $ where $\chi $ is an arbitrary real function of
time-space variables ($\chi =\chi \left( t,x_1,x_2,x_3\right) $):
\begin{eqnarray*}
&&\widetilde{U}^{\dagger }\left( \chi \right) \left( \frac 1{\mathrm{c}%
}\,\partial _t+\mathrm{i}\Theta _0+\mathrm{i}\Upsilon _0\gamma ^{\left[
5\right] }\right) \widetilde{U}\left( \chi \right) \varphi =\\
&&=\widetilde{U}^{\dagger }\left( \chi \right) \left( \sum\limits_{\nu
=1}^3\beta ^{\left[ \nu \right] }\left( \partial _\nu +\mathrm{i}\Theta _\nu
+\mathrm{i}\Upsilon _\nu \gamma ^{\left[ 5\right] }\right) \,+\widehat{M}%
\right) \widetilde{U}\left( \chi \right) \varphi \,.
\end{eqnarray*}

Because
\vspace*{-3pt}
\begin{eqnarray*}
&
\gamma ^{\left[ 5\right] }\widetilde{U}\left( \chi \right) =\widetilde{U}%
\left( \chi \right) \gamma ^{\left[ 5\right] }\,,
\\[8pt]
&
\beta ^{\left[ 1\right] }\widetilde{U}\left( \chi \right) =\widetilde{U}%
\left( \chi \right) \beta ^{\left[ 1\right] }\,,
\\[2pt]
&
\beta ^{\left[ 2\right] }\widetilde{U}\left( \chi \right) =\widetilde{U}%
\left( \chi \right) \beta ^{\left[ 2\right] }\,,
\\[2pt]
&
\beta ^{\left[ 3\right] }\widetilde{U}\left( \chi \right) =\widetilde{U}%
\left( \chi \right) \beta ^{\left[ 3\right] }\,,
\\[8pt]
&
\widetilde{U}^{\dagger }\left( \chi \right) \widetilde{U}\left( \chi \right)
=1_4\,,
\end{eqnarray*}

\vspace*{-6pt}\noindent
then
\begin{eqnarray*}
&&\rule{-1.3cm}{0pt}\left( \frac 1{\mathrm{c}}\,\partial _t+\frac 1{\mathrm{c%
}}\,\widetilde{U}^{\dagger }\left( \chi \right) \left( \partial _t\widetilde{%
U}\left( \chi \right) \right) +\mathrm{i}\Theta _0+\mathrm{i}\Upsilon
_0\gamma ^{\left[ 5\right] }\right) \varphi = \\
&&\rule{-1.3cm}{0pt}=\left( 
\begin{array}{c}
\sum\limits_{\nu =1}^3\beta ^{\left[ \nu \right] }\left( \partial _\nu +%
\widetilde{U}^{\dagger }\left( \chi \right) \left( \partial _\nu \widetilde{U%
}\left( \chi \right) \right) +\,\mathrm{i}\Theta _\nu +\mathrm{i}\Upsilon
_\nu \gamma ^{\left[ 5\right] }\right)  \\ 
+\,\widetilde{U}^{\dagger }\left( \chi \right) \widehat{M}\widetilde{U}%
\left( \chi \right) 
\end{array}
\right) \varphi \,.
\end{eqnarray*}

Now let:
\[
\widehat{U}\left( \kappa \right) :=
\left[ 
\begin{array}{cccc}
e^\kappa  & 0 & 0 & 0 \\ 
0 & e^\kappa  & 0 & 0 \\ 
0 & 0 & e^{2\kappa}  & 0 \\ 
0 & 0 & 0 & e^{2\kappa} 
\end{array}
\right] 
\]
and
\[
\widehat{M}^{\prime }:= 
\left(\begin{array}{c}
-\,M_{\zeta ,0}^{\prime }\gamma _\zeta ^{[0]}+M_{\zeta ,4}^{\prime }\zeta
^{[4]}- \\[3pt]
-M_{\eta ,0}^{\prime }\gamma _\eta ^{[0]}-M_{\eta ,4}^{\prime }\eta ^{[4]}+
\\[3pt]
+\,M_{\theta ,0}^{\prime }\gamma _\theta ^{[0]}+M_{\theta ,4}^{\prime }\theta
^{[4]}
\end{array}\right)
:=\widehat{U}^{-1}\left( \kappa \right) \widehat{M}\widehat{U}%
\left( \kappa \right) 
\]
then:
\vspace*{-6pt}
\begin{eqnarray*}
M_{\theta ,0}^{\prime } \!\!&\!\!{=}\!\!&\!\!\left( M_{\theta ,0}\cosh \kappa -\mathrm{i}%
M_{\theta ,4}\sinh \kappa \right) \mbox{,} \\[3pt]
M_{\theta ,4}^{\prime } \!\!&\!\!{=}\!\!&\!\!\left( M_{\theta ,4}\cosh \kappa +\mathrm{i}%
M_{\theta ,0}\sinh \kappa \right) \mbox{,} \\[3pt]
M_{\eta ,0}^{\prime } \!\!&\!\!{=}\!\!&\!\!\left( M_{\eta ,0}\cosh \kappa -\mathrm{i}M_{\eta
,4}\sinh \kappa \right) \mbox{,} \\[3pt]
M_{\eta ,4}^{\prime } \!\!&\!\!{=}\!\!&\!\!\left( M_{\eta ,4}\cosh \kappa +\mathrm{i}M_{\eta
,0}\sinh \kappa \right) \mbox{,} \\[3pt]
M_{\zeta ,0}^{\prime } \!\!&\!\!{=}\!\!&\!\!\left( M_{\zeta ,0}\cosh \kappa +\mathrm{i}%
M_{\zeta ,4}\sinh \kappa \right) \mbox{,} \\[3pt]
M_{\zeta ,4}^{\prime } \!\!&\!\!{=}\!\!&\!\!\left( M_{\zeta ,4}\cosh \kappa -\mathrm{i}%
M_{\zeta ,0}\sinh \kappa \right) \mbox{.}
\end{eqnarray*}

Therefore, matrix $\widehat{U}\left( \kappa \right) $ makes an oscillation
between upper and lower mass members, too.

Let us consider equation (\ref{clrH}) under transformation $\widehat{U}%
\left( \kappa \right) $ where $\kappa $ is an arbitrary real function of
time-space variables ($\kappa =\kappa \left( t,x_1,x_2,x_3\right) $):
\begin{eqnarray*}
&&\rule{-0.5cm}{0pt}\widehat{U}^{-1}\left( \kappa \right) \left( \frac 1{%
\mathrm{c}}\,\partial _t+\mathrm{i}\Theta _0+\mathrm{i}\Upsilon _0\gamma
^{\left[ 5\right] }\right) \widehat{U}\left( \kappa \right) \varphi = \\
&&\rule{-0.5cm}{0pt}=\widehat{U}^{-1}\left( \kappa \right) \left(
\sum\limits_{\nu =1}^3\beta ^{\left[ \nu \right] }\left( \partial _\nu +%
\mathrm{i}\Theta _\nu +\mathrm{i}\Upsilon _\nu \gamma ^{\left[ 5\right]
}\right) +\,\widehat{M}\right) \widehat{U}\left( \kappa \right) \varphi\mbox{.} 
\end{eqnarray*}

Because
\begin{eqnarray*}
&
\gamma ^{\left[ 5\right] }\widehat{U}\left( \kappa \right) =\widehat{U}%
\left( \kappa \right) \gamma ^{\left[ 5\right] }\,,
\\[8pt]
&
\widehat{U}^{-1}\left( \kappa \right) \beta ^{\left[ 1\right] } =\beta
^{\left[ 1\right] }\widehat{U}^{-1}\left( \kappa \right)\,,
\\[2pt]
&
\widehat{U}^{-1}\left( \kappa \right) \beta ^{\left[ 2\right] } =\beta
^{\left[ 2\right] }\widehat{U}^{-1}\left( \kappa \right),
\\[2pt]
&
\widehat{U}^{-1}\left( \kappa \right) \beta ^{\left[ 3\right] } =\beta
^{\left[ 3\right] }\widehat{U}^{-1}\left( \kappa \right),
\\[8pt]
&
\widehat{U}^{-1}\left( \kappa \right) \widehat{U}\left( \kappa \right) =1_4\,,
\end{eqnarray*}
then
\begin{eqnarray*}
&&\rule{-0.5cm}{0pt}\left( \frac 1{\mathrm{c}}\,\partial _t+\widehat{U}%
^{-1}\left( \kappa \right) \left( \frac 1{\mathrm{c}}\,\partial _t\widehat{U}%
\left( \kappa \right) \right) +\mathrm{i}\Theta _0+\mathrm{i}\Upsilon
_0\gamma ^{\left[ 5\right] }\right) \varphi = \\[8pt]
&&\rule{-0.5cm}{0pt}=\left( 
\begin{array}{c}
\sum\limits_{\nu =1}^3\beta ^{\left[ \nu \right] }\left( \partial _\nu +%
\widehat{U}^{-1}\left( \kappa \right) \left( \partial _\nu \widehat{U}\left(
\kappa \right) \right) +\,\mathrm{i}\Theta _\nu +\mathrm{i}\Upsilon _\nu
\gamma ^{\left[ 5\right] }\right) + \\[15pt]
+\,\widehat{U}^{-1}\left( \kappa \right) \widehat{M}\widehat{U}\left( \kappa
\right) 
\end{array}
\right) \varphi \,.
\end{eqnarray*}

If denote:
\[
\ \Lambda _1:=\left[ 
\begin{array}{cccc}
0 & -1 & 0 & 0 \\ 
-1 & 0 & 0 & 0 \\ 
0 & 0 & 0 & 1 \\ 
0 & 0 & 1 & 0
\end{array}
\right] \mbox{, }
\]
\[
\Lambda _2:=\left[ 
\begin{array}{cccc}
0 & \mathrm{i} & 0 & 0 \\ 
\mathrm{i} & 0 & 0 & 0 \\ 
0 & 0 & 0 & \mathrm{i} \\ 
0 & 0 & \mathrm{i} & 0
\end{array}
\right] \mbox{, }
\]
\[
\ \Lambda _3:=\left[ 
\begin{array}{cccc}
0 & 1 & 0 & 0 \\ 
-1 & 0 & 0 & 0 \\ 
0 & 0 & 0 & 1 \\ 
0 & 0 & -1 & 0
\end{array}
\right] \mbox{, }
\]
\[
\Lambda _4:=\left[ 
\begin{array}{cccc}
0 & \mathrm{i} & 0 & 0 \\ 
-\mathrm{i} & 0 & 0 & 0 \\ 
0 & 0 & 0 & -\mathrm{i} \\ 
0 & 0 & \mathrm{i} & 0
\end{array}
\right] \mbox{,}
\]
\[
\ \Lambda _5:=\left[ 
\begin{array}{cccc}
-\mathrm{i} & 0 & 0 & 0 \\ 
0 & \mathrm{i} & 0 & 0 \\ 
0 & 0 & -\mathrm{i} & 0 \\ 
0 & 0 & 0 & \mathrm{i}
\end{array}
\right] \mbox{, }
\]
\[
\Lambda _6:=\left[ 
\begin{array}{cccc}
1 & 0 & 0 & 0 \\ 
0 & -1 & 0 & 0 \\ 
0 & 0 & -1 & 0 \\ 
0 & 0 & 0 & 1
\end{array}
\right] \mbox{,}
\]
\[
\Lambda _7:=\left[ 
\begin{array}{cccc}
1 & 0 & 0 & 0 \\ 
0 & 1 & 0 & 0 \\ 
0 & 0 & 2 & 0 \\ 
0 & 0 & 0 & 2
\end{array}
\right] \mbox{, }
\]
\[
\Lambda _8:=\left[ 
\begin{array}{cccc}
\mathrm{i} & 0 & 0 & 0 \\ 
0 & \mathrm{i} & 0 & 0 \\ 
0 & 0 & 2\mathrm{i} & 0 \\ 
0 & 0 & 0 & 2\mathrm{i}
\end{array}
\right] \mbox{,}
\]
then
\begin{eqnarray*}
&&\rule{-.5cm}{0pt}
U_{0,1}^{-1}\left( \sigma \right) \bigl( \partial _sU_{0,1}\left( \sigma
\right) \bigr) =\Lambda _1\partial _s\sigma \,\mbox{,} \\[3pt]
&&\rule{-.5cm}{0pt}
U_{2,3}^{-1}\left( \alpha \right) \bigl( \partial _sU_{2,3}\left( \alpha
\right) \bigr) =\Lambda _2\partial _s\alpha \,\mbox{,} \\[3pt]
&&\rule{-.5cm}{0pt}
U_{1,3}^{-1}\left( \vartheta \right) \bigl( \partial _sU_{1,3}\left(
\vartheta \right) \bigr) =\Lambda _3\partial _s\vartheta \,\mbox{,} \\[3pt]
&&\rule{-.5cm}{0pt}
U_{0,2}^{-1}\left( \phi \right) \bigl( \partial _sU_{0,2}\left( \phi \right)
\bigr) =\Lambda _4\partial _s\phi \,\mbox{,} \\[3pt]
&&\rule{-.5cm}{0pt}
U_{1,2}^{-1}\left( \varsigma \right) \bigl( \partial _sU_{1,2}\left(
\varsigma \right) \bigr) =\Lambda _5\partial _s\varsigma \,\mbox{,} \\[3pt]
&&\rule{-.5cm}{0pt}
U_{0,3}^{-1}\left( \iota \right) \bigl( \partial _sU_{0,3}\left( \iota
\right) \bigr) =\Lambda _6\partial _s\iota \,\mbox{,} \\[3pt]
&&\rule{-.5cm}{0pt}
\widehat{U}^{-1}\left( \kappa \right) \bigl( \partial _s\widehat{U}\left(
\kappa \right) \bigr) =\Lambda _7\partial _s\kappa \,\mbox{,} \\[3pt]
&&\rule{-.5cm}{0pt}
\widetilde{U}^{-1}\left( \chi \right) \bigl( \partial _s\widetilde{U}\left(
\chi \right) \bigr) =\Lambda _8\partial _s\chi \,\mbox{.}
\end{eqnarray*}

Let $\grave U$ be the following set:
\[
\grave U:=\left\{
U_{0,1},U_{2,3},U_{1,3},U_{0,2},U_{1,2},U_{0,3},\widehat{U},\widetilde{U}%
\right\} \mbox{.} 
\]

Because

\medskip\vspace*{2pt}
$U_{2,3}^{-1}\left( \alpha \right) \Lambda _1U_{2,3}\left( \alpha \right)
=\Lambda _1$

\smallskip
$U_{1,3}^{-1}\left( \vartheta \right) \Lambda _1U_{1,3}\left( \vartheta
\right) =\left( \Lambda _1\cos 2\vartheta +\Lambda _6\sin 2\vartheta \right) 
$

\smallskip
$U_{0,2}^{-1}\left( \phi \right) \Lambda _1U_{0,2}\left( \phi \right)
=\left( \Lambda _1\cosh 2\phi -\Lambda _5\sinh 2\phi \right) $

\smallskip
$U_{1,2}^{-1}\left( \varsigma \right) \Lambda _1U_{1,2}\left( \varsigma
\right) =\Lambda _1\cos 2\varsigma -\Lambda _4\sin 2\varsigma $

\smallskip
$U_{0,3}^{-1}\left( \iota \right) \Lambda _1U_{0,3}\left( \iota \right)
=\Lambda _1\cosh 2\iota +\Lambda _3\sinh 2\iota $

\smallskip
$\widehat{U}^{-1}\left( \kappa \right) \Lambda _1\widehat{U}\left( \kappa
\right) =\Lambda _1$

\smallskip
$\widetilde{U}^{-1}\left( \chi \right) \Lambda _1\widetilde{U}\left( \chi
\right) =\Lambda _1$

\smallskip
========

\smallskip
$\widetilde{U}^{-1}\left( \chi \right) \Lambda _2\widetilde{U}\left( \chi
\right) =\Lambda _2$

\smallskip
$\widehat{U}^{-1}\left( \kappa \right) \Lambda _2\widehat{U}\left( \kappa
\right) =\Lambda _2$

\smallskip
$U_{0,3}^{-1}\left( \iota \right) \Lambda _2U_{0,3}\left( \iota \right)
=\Lambda _2\cosh 2\iota -\Lambda _4\sinh 2\iota $

\smallskip
$U_{1,2}^{-1}\left( \varsigma \right) \Lambda _2U_{1,2}\left( \varsigma
\right) =\Lambda _2\cos 2\varsigma -\Lambda _3\sin 2\varsigma $

\smallskip
$U_{0,2}^{-1}\left( \phi \right) \Lambda _2U_{0,2}\left( \phi \right)
=\Lambda _2\cosh 2\phi +\Lambda _6\sinh 2\phi $

\smallskip
$U_{1,3}^{-1}\left( \vartheta \right) \Lambda _2U_{1,3}\left( \vartheta
\right) =\Lambda _2\cos 2\vartheta +\Lambda _5\sin 2\vartheta $

\smallskip
$U_{0,1}^{-1}\left( \sigma \right) \Lambda _2U_{0,1}\left( \sigma
\right) =\Lambda _2$

\smallskip
=========

\smallskip
$U_{0,1}^{-1}\left( \sigma \right) \ \Lambda _3U_{0,1}\left( \sigma
\right) =\Lambda _3\cosh 2\sigma -\Lambda _6\sinh 2\sigma $

\smallskip
$U_{2,3}^{-1}\left( \alpha \right) \ \Lambda _3U_{2,3}\left( \alpha \right)
=\Lambda _3\cos 2\alpha -\Lambda _5\sin 2\alpha $

\smallskip
$U_{0,2}^{-1}\left( \phi \right) \ \Lambda _3U_{0,2}\left( \phi \right)
=\Lambda _3$

\smallskip
$U_{1,2}^{-1}\left( \varsigma \right) \ \Lambda _3U_{1,2}\left( \varsigma
\right) =\Lambda _3\cos 2\varsigma +\Lambda _2\sin 2\varsigma $

\smallskip
$U_{0,3}^{-1}\left( \iota \right) \ \Lambda _3U_{0,3}\left( \iota \right) =\
\Lambda _3\cosh 2\iota +\Lambda _1\sinh 2\iota $

\smallskip
$\widehat{U}^{-1}\left( \kappa \right) \ \Lambda _3\widehat{U}\left( \kappa
\right) =\Lambda _3$

\smallskip
$\widetilde{U}^{-1}\left( \chi \right) \ \Lambda _3\widetilde{U}\left( \chi
\right) =\Lambda _3$

\smallskip
==========

\smallskip
$\widetilde{U}^{-1}\left( \chi \right) \ \Lambda _4\widetilde{U}\left( \chi
\right) =\Lambda _4$

\smallskip
$\widehat{U}^{-1}\left( \kappa \right) \ \Lambda _4\widehat{U}\left( \kappa
\right) =\Lambda _4$

\smallskip
$U_{0,3}^{-1}\left( \iota \right) \ \Lambda _4U_{0,3}\left( \iota \right)
=\Lambda _4\cosh 2\iota -\Lambda _2\sinh 2\iota $

\smallskip
$U_{1,2}^{-1}\left( \varsigma \right) \ \Lambda _4U_{1,2}\left( \varsigma
\right) =\Lambda _4\cos 2\varsigma +\Lambda _1\sin 2\varsigma $

\smallskip
$U_{1,3}^{-1}\left( \vartheta \right) \ \Lambda _4U_{1,3}\left( \vartheta
\right) =\Lambda _4$

\smallskip
$U_{2,3}^{-1}\left( \alpha \right) \ \Lambda _4U_{2,3}\left( \alpha \right)
=\Lambda _4\cos 2\alpha -\Lambda _6\sin 2\alpha $

\smallskip
$U_{0,1}^{-1}\left( \sigma \right) \ \Lambda _4U_{0,1}\left( \sigma
\right) =\Lambda _4\cosh 2\sigma +\Lambda _5\sinh 2\sigma $

\smallskip
==========

\smallskip
$U_{0,1}^{-1}\left( \sigma \right) \ \Lambda _5U_{0,1}\left( \sigma
\right) \ =\Lambda _5\cosh 2\sigma +\Lambda _4\sinh 2\sigma $

\smallskip
$U_{2,3}^{-1}\left( \alpha \right) \ \Lambda _5U_{2,3}\left( \alpha \right)
\ =\Lambda _5\cos 2\alpha +\Lambda _3\sin 2\alpha $

\smallskip
$U_{1,3}^{-1}\left( \vartheta \right) \ \Lambda _5U_{1,3}\left( \vartheta
\right) \ =\left( \Lambda _5\cos 2\vartheta -\Lambda _2\sin 2\vartheta
\right) $

\smallskip
$U_{0,2}^{-1}\left( \phi \right) \ \Lambda _5U_{0,2}\left( \phi \right) \
=\Lambda _5\cosh 2\phi -\Lambda _1\sinh 2\phi $

\smallskip
$U_{0,3}^{-1}\left( \iota \right) \ \Lambda _5U_{0,3}\left( \iota \right) \
=\Lambda _5$

\smallskip
$\widehat{U}^{-1}\left( \kappa \right) \ \Lambda _5\widehat{U}\left( \kappa
\right) \ =\Lambda _5$

\smallskip
$\widetilde{U}^{-1}\left( \chi \right) \Lambda _5\widetilde{U}\left( \chi
\right) \ =\Lambda _5$

\smallskip
===========

\smallskip
$\widetilde{U}^{-1}\left( \chi \right) \Lambda _6\widetilde{U}\left( \chi
\right) \ =\Lambda _6$

\smallskip
$\widehat{U}^{-1}\left( \kappa \right) \Lambda _6\widehat{U}\left( \kappa
\right) \ =\Lambda _6$

\smallskip
$U_{1,2}^{-1}\left( \varsigma \right) \Lambda _6U_{1,2}\left( \varsigma
\right) \ =\Lambda _6$

\smallskip
$U_{0,2}^{-1}\left( \phi \right) \Lambda _6U_{0,2}\left( \phi \right) \
=\Lambda _6\cosh 2\phi +\Lambda _2\sinh 2\phi $

\smallskip
$U_{1,3}^{-1}\left( \vartheta \right) \Lambda _6U_{1,3}\left( \vartheta
\right) \ =\Lambda _6\cos 2\vartheta -\Lambda _1\sin 2\vartheta $

\smallskip
$U_{2,3}^{-1}\left( \alpha \right) \Lambda _6U_{2,3}\left( \alpha \right) \
=\Lambda _6\cos 2\alpha +\Lambda _4\sin 2\alpha $

\smallskip
$U_{0,1}^{-1}\left( \sigma \right) \Lambda _6U_{0,1}\left( \sigma
\right) =\Lambda _6\cosh 2\sigma -\Lambda _3\sinh 2\sigma $

\smallskip
========

\smallskip
$\widetilde{U}^{-1}\left( \chi \right) \ \Lambda _7\widetilde{U}\left( \chi
\right) =\Lambda _7$

\smallskip
$U_{0,3}^{-1}\left( \iota \right) \ \Lambda _7U_{0,3}\left( \iota \right)
=\Lambda _7$

\smallskip
$U_{1,2}^{-1}\left( \varsigma \right) \ \Lambda _7U_{1,2}\left( \varsigma
\right) =\Lambda _7$

\smallskip
$U_{0,2}^{-1}\left( \phi \right) \ \Lambda _7U_{0,2}\left( \phi \right)
=\Lambda _7$

\smallskip
$U_{1,3}^{-1}\left( \vartheta \right) \ \Lambda _7U_{1,3}\left( \vartheta
\right) =\Lambda _7$

\smallskip
$U_{2,3}^{-1}\left( \alpha \right) \ \Lambda _7U_{2,3}\left( \sigma
\right) =\Lambda _7$

\smallskip
$U_{0,1}^{-1}\left( \sigma \right) \ \Lambda _7U_{0,1}\left( \sigma
\right) =\Lambda _7$

\smallskip
=========

\smallskip
$U_{0,1}^{-1}\left( \sigma \right) \ \Lambda _8U_{0,1}\left( \sigma
\right) =\Lambda _8$

\smallskip
$U_{2,3}^{-1}\left( \alpha \right) \ \Lambda _8U_{2,3}\left( \alpha \right)
=\Lambda _8$

\smallskip
$U_{1,3}^{-1}\left( \vartheta \right) \ \Lambda _8U_{1,3}\left( \vartheta
\right) =\Lambda _8$

\smallskip
$U_{0,2}^{-1}\left( \phi \right) \ \Lambda _8U_{0,2}\left( \phi \right)
=\Lambda _8$

\smallskip
$U_{1,2}^{-1}\left( \varsigma \right) \ \Lambda _8U_{1,2}\left( \varsigma
\right) =\Lambda _8$

\smallskip
$U_{0,3}^{-1}\left( \iota \right) \ \Lambda _8U_{0,3}\left( \iota \right)
=\Lambda _8$

\smallskip
$\widehat{U}^{-1}\left( \kappa \right) \ \Lambda _8\widehat{U}\left( \kappa
\right) =\Lambda _8$

\medskip\noindent
then for every product $U$ of $\grave U$'s elements real
functions $G_s^r\left( t,x_1,x_2,x_3\right) $ exist such that
\vspace*{-2pt}
\[
U^{-1}\left( \partial _sU\right) =\frac{g_3}2\sum_{r=1}^8\Lambda _rG_s^r
\]

\vspace*{-2pt}\noindent
with some real constant $g_3$ (similar to 8 gluons). 

\subsection{Asymptotic Freedom, Confinement, Gravitation} 

From (\ref{grg}):

\begin{eqnarray}
\frac{\partial t}{\partial t^{\prime }} &=&\cosh 2\sigma \mbox{,}  \label{1}
\\
\frac{\partial x}{\partial t^{\prime }} &=&c\sinh 2\sigma \mbox{.}  \nonumber
\end{eqnarray}

Hence, if $v$ is the velocity of a coordinate system $\left\{ t^{\prime },x^{\prime }\right\} $ in 
the coordinate system $\left\{t,x\right\} $  then

\[
\sinh 2\sigma =\frac{\left( \frac v{\mathrm{c}}\right) }{\sqrt{1-\left(
\frac v{\mathrm{c}}\right) ^2}}\mbox{, }\cosh 2\sigma =\frac 1{\sqrt{%
1-\left( \frac v{\mathrm{c}}\right) ^2}}\mbox{.}
\]

Therefore,

\begin{equation}
v=\mathrm{c}\tanh 2\sigma\mbox{.}   \label{2}
\end{equation}

Let 
\[
2\sigma :=\omega \left( x\right) \frac t{x}
\]

\vspace*{-4pt}\noindent
with
\[
\omega \left( x\right) =\frac \lambda {\left| x\right| }\,,
\]
where $\lambda$ is a real constant with positive numerical value. 

In that case
 
\begin{equation}
v\left( t,x\right) =\mathrm{c}\tanh \left( \frac \lambda {\left| x\right|
}\frac tx\right) \mbox{.}  \label{3}
\end{equation}

and if $\mathrm{g}$ is an acceleration of system $\left\{ t^{\prime
},x_1^{\prime }\right\} $ as respects to system $\left\{ t,x_1\right\} $ then
\vspace*{-2pt}
\begin{equation}
\mathrm{g}\left( t,x_1\right) =\frac{\partial v}{\partial t}=\frac{\mathrm{c}\omega
\left( x_1\right) }{\left( \cosh ^2\omega \left( x_1\right) \frac t{x_1}\right)
x_1}\,\mbox{.}  \label{acs}
\end{equation}

\begin{figure}[htbp]
\centering
\includegraphics[width=.75\textwidth]{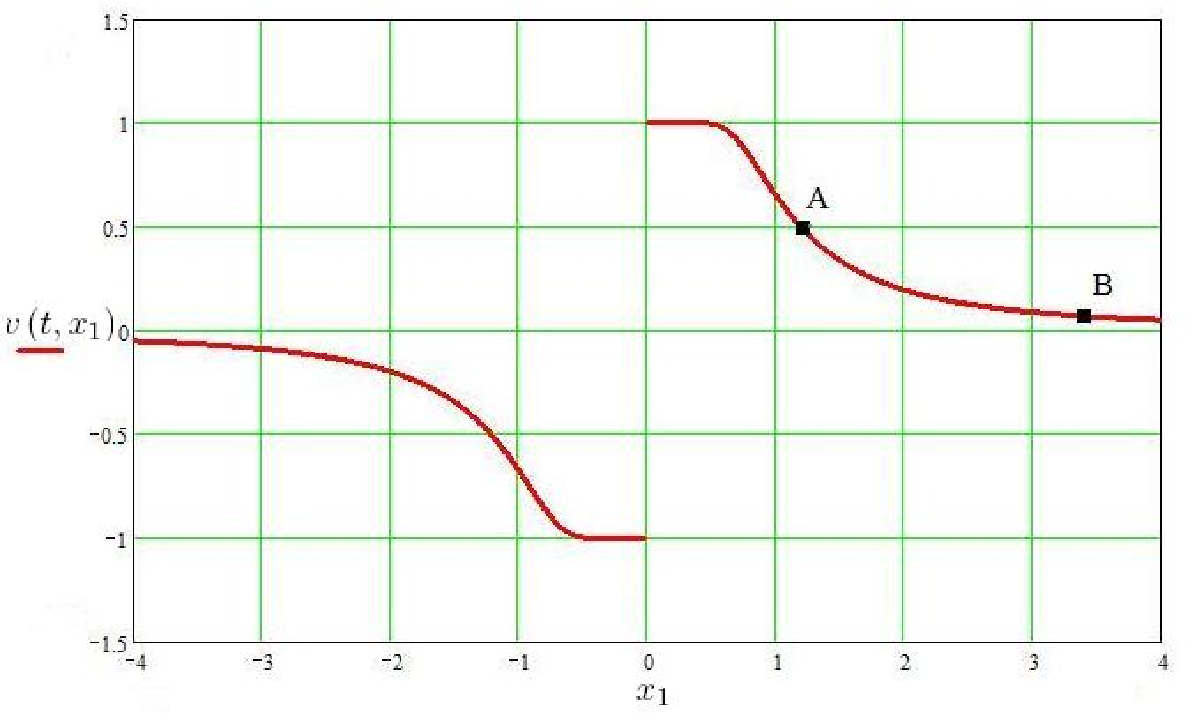}
\caption{the dependency of a system $\left\{ t^{\prime },x_1^{\prime}\right\} $ velocity $v\left( t,x_1\right) $ 
on $x_1$ in system $\left\{ t,x_1\right\} $.}
\end{figure}

\begin{figure}[htbp]
\centering
\includegraphics[width=.75\textwidth]{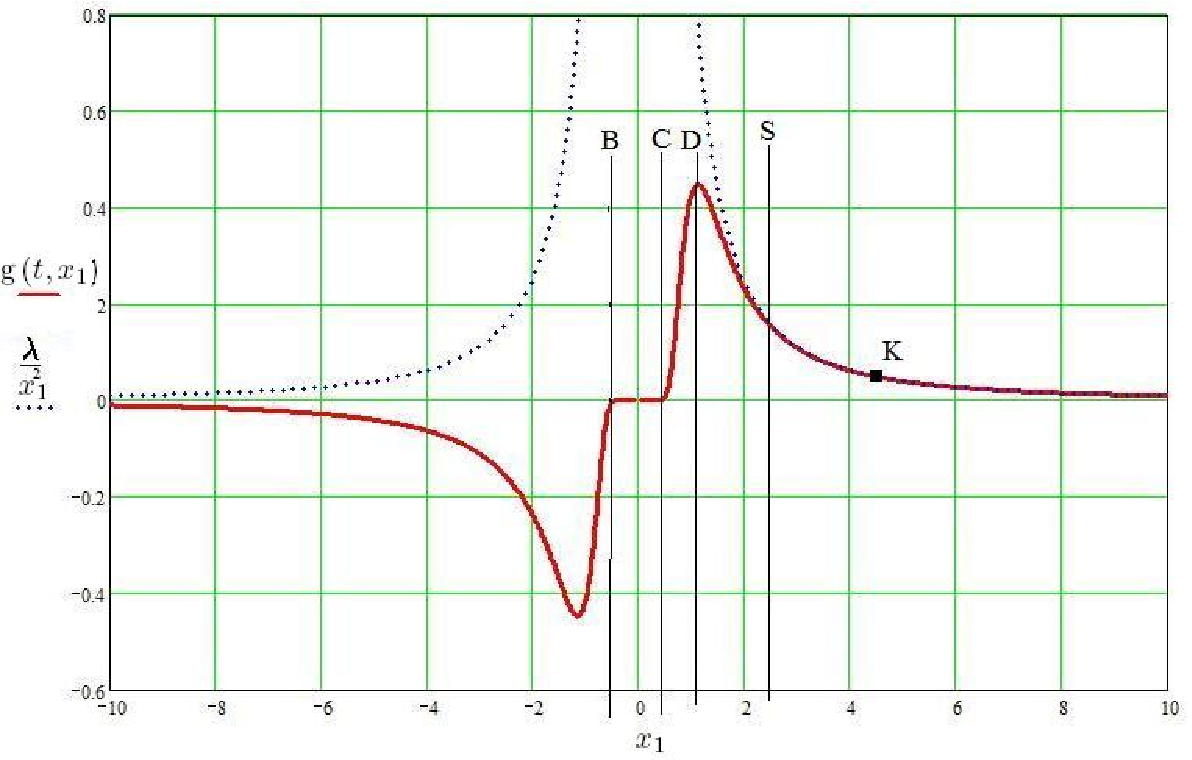}
\caption{the dependency of a system $\left\{ t^{\prime },x_1^{\prime
}\right\} $ acceleration $g\left( t,x_1\right) $ on $x_1$ in system $\left\{ t,x_1\right\} $.}
\end{figure}

\vspace*{-3pt}
Figure 1 shows the dependency of a system $\left\{ t^{\prime },x_1^{\prime}\right\} $ velocity $v\left( t,x_1\right) $ 
on $x_1$ in system $\left\{ t,x_1\right\} $.

This velocity in point $A$ is not equal to one in point $B$. Hence, an oscillator, 
placed in $B$, has a nonzero velocity in respect to an observer, placed in point 
$A$. Therefore, from the Lorentz transformations, this oscillator frequency for observer,
placed in point $A$, is less than own frequency of this oscillator ({\it red shift}).

Figure 2 shows the dependency of a system $\left\{ t^{\prime },x_1^{\prime
}\right\} $ acceleration $g\left( t,x_1\right) $ on $x_1$ in system $\left\{ t,x_1\right\} $.

If an object immovable in system $\left\{ t,x_1\right\} $ is placed in point $K$ then 
in system $\left\{ t^{\prime },x_1^{\prime}\right\} $ this object must move to the left 
with acceleration $g$ and $\mathrm{g}\simeq \frac \lambda {x_1^2}$. 

I call:
\begin{itemize}
\item 
interval from $S$ to $\infty $ the {\it Newton Gravity Zone},
\item 
interval from $B$ to $C$ the {\it the Zone},
\item 
and interval from $C$ to $D$ {\it the Confinement Force Zone}.
\end{itemize}

\subsection{Baryon Chrome}

Like coordinates $x_5$ and $x_4$ \cite{Q1} here are entered new
coordinates $y^\beta $, $z^\beta $, $y^\zeta $, $z^\zeta $, $y^\eta $, $%
z^\eta $, $y^\theta $, $z^\theta $ such that

\begin{eqnarray*}
-\frac{\pi \mathrm{c}}{\mathrm{h}} &\leq &y^\beta \leq \frac{\pi \mathrm{c}}{%
\mathrm{h}}\mbox{,}-\frac{\pi \mathrm{c}}{\mathrm{h}}\leq z^\beta \leq \frac{%
\pi \mathrm{c}}{\mathrm{h}}\mbox{,} \\
-\frac{\pi \mathrm{c}}{\mathrm{h}} &\leq &y^\zeta \leq \frac{\pi \mathrm{c}}{%
\mathrm{h}}\mbox{,}-\frac{\pi \mathrm{c}}{\mathrm{h}}\leq z^\zeta \leq \frac{%
\pi \mathrm{c}}{\mathrm{h}}\mbox{,} \\
-\frac{\pi \mathrm{c}}{\mathrm{h}} &\leq &y^\eta \leq \frac{\pi \mathrm{c}}{%
\mathrm{h}}\mbox{,}-\frac{\pi \mathrm{c}}{\mathrm{h}}\leq z^\eta \leq \frac{%
\pi \mathrm{c}}{\mathrm{h}}\mbox{,} \\
-\frac{\pi \mathrm{c}}{\mathrm{h}} &\leq &y^\theta \leq \frac{\pi \mathrm{c}%
}{\mathrm{h}}\mbox{,}-\frac{\pi \mathrm{c}}{\mathrm{h}}\leq z^\theta \leq 
\frac{\pi \mathrm{c}}{\mathrm{h}}\mbox{.}
\end{eqnarray*}

\noindent
and like $\widetilde{\varphi }$, \cite[p.83]{Q1} let:

\begin{eqnarray}
&&\ \left[ \varphi \right] \left( t,\mathbf{x},y^\beta ,z^\beta ,y^\zeta
,z^\zeta ,y^\eta ,z^\eta ,y^\theta ,z^\theta \right)  :=  \label{fff} \\
\  &:&=\varphi \left( t,\mathbf{x}\right) \times \exp (\mathrm{i}(y^\beta
M_0+z^\beta M_4+y^\zeta M_{\zeta ,0}+z^\zeta M_{\zeta ,4}+  \nonumber \\
&&\ +y^\eta M_{\eta ,0}+z^\eta M_{\eta ,4}+y^\theta M_{\theta ,0}+z^\theta
M_{\theta ,4}))\mbox{,}  \nonumber
\end{eqnarray}

In fhis case if 
\begin{equation}
\begin{array}{c}
\displaystyle\left( \left[ \varphi \right] ,\left[ \chi \right] \right) :=
\\ 
\displaystyle:=\int_{-\frac{\pi \mathrm{c}}{h}}^{\frac{\pi \mathrm{c}}{%
\mathrm{h}}}dy^{\beta }\int_{-\frac{\pi \mathrm{c}}{\mathrm{h}}}^{\frac{\pi 
\mathrm{c}}{\mathrm{h}}}dz^{\beta }\int_{-\frac{\pi \mathrm{c}}{h}}^{\frac{%
\pi \mathrm{c}}{\mathrm{h}}}dy^{\zeta }\int_{-\frac{\pi \mathrm{c}}{\mathrm{h%
}}}^{\frac{\pi \mathrm{c}}{\mathrm{h}}}dz^{\zeta }\times  \\[12pt]
\displaystyle\times \int_{-\frac{\pi \mathrm{c}}{h}}^{\frac{\pi \mathrm{c}}{%
\mathrm{h}}}dy^{\eta }\int_{-\frac{\pi \mathrm{c}}{\mathrm{h}}}^{\frac{\pi 
\mathrm{c}}{\mathrm{h}}}dz^{\eta }\int_{-\frac{\pi \mathrm{c}}{h}}^{\frac{%
\pi \mathrm{c}}{\mathrm{h}}}dy^{\theta }\int_{-\frac{\pi \mathrm{c}}{\mathrm{%
h}}}^{\frac{\pi \mathrm{c}}{\mathrm{h}}}dz^{\theta }\times  \\[12pt]
\displaystyle\times \quad \left[ \varphi \right] ^{\dagger }\left[ \chi %
\right] 
\end{array}
\label{sc}
\end{equation}%
then 
\begin{eqnarray}
\left( \left[ \varphi \right] ,\left[ \varphi \right] \right)  &=&\rho _{%
\mathcal{A}}\mbox{,}  \label{jaxx} \\
\left( \left[ \varphi \right] ,\beta ^{\left[ s\right] }\left[ \varphi %
\right] \right)  &=&-\frac{j_{\mathcal{A},k}}{\mathrm{c}}\mbox{.}  \nonumber
\end{eqnarray}
and in this case from (\ref{clrH}):
\begin{equation}
\begin{array}{c}
(\sum\limits_{\nu =0}^{3}\beta ^{\left[ \nu \right] }\left( \partial _{\nu }+%
\mathrm{i}\Theta _{\nu }+\mathrm{i}\Upsilon _{\nu }\gamma ^{\left[ 5\right]
}\right) + \\ 
+\gamma ^{\left[ 0\right] }\partial _{y}^{\beta }+\beta ^{\left[ 4\right]
}\partial _{z}^{\beta }- \\ 
-\gamma _{\zeta }^{[0]}\partial _{y}^{\zeta }+\zeta ^{\lbrack 4]}\partial
_{z}^{\zeta }- \\ 
-\gamma _{\eta }^{[0]}\partial _{y}^{\eta }-\eta ^{\lbrack 4]}\partial
_{z}^{\eta }+ \\ 
+\gamma _{\theta }^{[0]}\partial _{y}^{\theta }+\theta ^{\lbrack 4]}\partial
_{z}^{\theta })\left[ \varphi \right] =0%
\end{array}%
\mbox{.}  \label{ff100}
\end{equation}

Because

\begin{equation}
\gamma _\eta ^{[0]}=\left[ 
\begin{array}{cccc}
0 & 0 & 0 & i \\ 
0 & 0 & -i & 0 \\ 
0 & i & 0 & 0 \\ 
-i & 0 & 0 & 0
\end{array}
\right] \mbox{, }\eta ^{[4]}=\mathrm{i}\left[ 
\begin{array}{cccc}
0 & 0 & 0 & -i \\ 
0 & 0 & i & 0 \\ 
0 & i & 0 & 0 \\ 
-i & 0 & 0 & 0
\end{array}
\right] \mbox{;}  \label{greenMM}
\end{equation}

\begin{equation}
\gamma _\theta ^{[0]}=\left[ 
\begin{array}{cccc}
0 & 0 & -1 & 0 \\ 
0 & 0 & 0 & 1 \\ 
-1 & 0 & 0 & 0 \\ 
0 & 1 & 0 & 0
\end{array}
\right] \mbox{, }\theta ^{[4]}=\mathrm{i}\left[ 
\begin{array}{cccc}
0 & 0 & 1 & 0 \\ 
0 & 0 & 0 & -1 \\ 
-1 & 0 & 0 & 0 \\ 
0 & 1 & 0 & 0
\end{array}
\right]   \label{blueMm}
\end{equation}

\begin{equation}
\gamma _\zeta ^{[0]}=\left[ 
\begin{array}{cccc}
0 & 0 & 0 & -1 \\ 
0 & 0 & -1 & 0 \\ 
0 & -1 & 0 & 0 \\ 
-1 & 0 & 0 & 0
\end{array}
\right] \mbox{, }\zeta ^{[4]}=\mathrm{i}\left[ 
\begin{array}{cccc}
0 & 0 & 0 & 1 \\ 
0 & 0 & 1 & 0 \\ 
0 & -1 & 0 & 0 \\ 
-1 & 0 & 0 & 0
\end{array}
\right] \mbox{;}  \label{redMM}
\end{equation}

then from (\ref{ff100}):
\begin{equation}
\begin{array}{c}
\sum\limits_{\nu =0}^{3}\beta ^{\left[ \nu \right] }\left( \partial _{\nu }+%
\mathrm{i}\Theta _{\nu }+\mathrm{i}\Upsilon _{\nu }\gamma ^{\left[ 5\right]
}\right) \left[ \varphi \right]  \\ 
+\gamma ^{\left[ 0\right] }\partial _{y}^{\beta }\left[ \varphi \right]
+\beta ^{\left[ 4\right] }\partial _{z}^{\beta }\left[ \varphi \right] + \\%
[12pt]
\begin{array}{c}
(\left[ 
\begin{array}{cccc}
0 & 0 & -\partial _{y}^{\theta } & \partial _{y}^{\zeta }-i\partial
_{y}^{\eta } \\ 
0 & 0 & \partial _{y}^{\zeta }+i\partial _{y}^{\eta } & \partial
_{y}^{\theta } \\ 
-\partial _{y}^{\theta } & \partial _{y}^{\zeta }-i\partial _{y}^{\eta } & 0
& 0 \\ 
\partial _{y}^{\zeta }+i\partial _{y}^{\eta } & \partial _{y}^{\theta } & 0
& 0%
\end{array}%
\right] + \\ 
\mathrm{i}\left[ 
\begin{array}{cccc}
0 & 0 & \partial _{z}^{\theta } & \partial _{z}^{\zeta }+i\partial
_{z}^{\eta } \\ 
0 & 0 & \partial _{z}^{\zeta }-i\partial _{z}^{\eta } & -\partial
_{z}^{\theta } \\ 
-\partial _{z}^{\theta } & -\partial _{z}^{\zeta }-i\partial _{z}^{\eta } & 0
& 0 \\ 
-\partial _{z}^{\zeta }+i\partial _{z}^{\eta } & \partial _{z}^{\theta } & 0
& 0%
\end{array}%
\right] )%
\end{array}
\\ 
\times \left[ \varphi \right] =0\mbox{.}%
\end{array}
\label{ffD}
\end{equation}

Let a Fourier transformation of 
\[
\left[ \varphi \right] \left( t,\mathbf{x}%
,y^{\beta },z^{\beta },y^{\zeta },z^{\zeta },y^{\eta },z^{\eta },y^{\theta
},z^{\theta }\right) 
\] 
be the following;

\begin{eqnarray}
&&\left[ \varphi \right] \left( t,\mathbf{x},y^{\beta },z^{\beta },y^{\zeta
},z^{\zeta },y^{\eta },z^{\eta },y^{\theta },z^{\theta }\right) =  \nonumber
\label{fun} \\
&=&\sum_{w,p_{1},p_{2},p_{3},n^{\beta },s^{\beta },n^{\zeta },s^{\zeta
},n^{\eta },s^{\eta },n^{\theta },s^{\theta }}c(w,p_{1},p_{2},p_{3},n^{\beta
},s^{\beta },  \nonumber \\
&&n^{\zeta },s^{\zeta },n^{\eta },s^{\eta },n^{\theta },s^{\theta })\times  
\nonumber \\
&&\times \exp (-\mathrm{i}\frac{\mathrm{h}}{\mathrm{c}}%
(wx_{0}+p_{1}x_{1}+p_{2}x_{2}+p_{3}x_{3}+ \\
&&+n^{\beta }y^{\beta }+s^{\beta }z^{\beta }\ +n^{\zeta }y^{\zeta }+s^{\zeta
}z^{\zeta }+  \nonumber \\
&&+n^{\eta }y^{\eta }+s^{\eta }z^{\eta }+n^{\theta }y^{\theta }+s^{\theta
}z^{\theta }))\mbox{.}  \nonumber
\end{eqnarray}

Let in (\ref{ffD}) $\Theta _{\nu }=0$ and $\Upsilon _{\nu }=0$.

Let us designe:

\begin{equation}
\begin{array}{c}
G_{0}:=(\sum\limits_{\nu =0}^{3}\beta ^{\left[ \nu \right] }\partial _{\nu
}+\gamma ^{\left[ 0\right] }\partial _{y}^{\beta }+\beta ^{\left[ 4\right]
}\partial _{z}^{\beta }- \\ 
-\gamma _{\zeta }^{[0]}\partial _{y}^{\zeta }+\zeta ^{\lbrack 4]}\partial
_{z}^{\zeta }- \\ 
-\gamma _{\eta }^{[0]}\partial _{y}^{\eta }-\eta ^{\lbrack 4]}\partial
_{z}^{\eta }+ \\ 
+\gamma _{\theta }^{[0]}\partial _{y}^{\theta }+\theta ^{\lbrack 4]}\partial
_{z}^{\theta })\mbox{.}%
\end{array}
\label{G0}
\end{equation}
               
\noindent
that is:
\begin{equation}
\rule{-1.8cm}{0pt}
\begin{array}{c}
G_{0}= \\ 
\left[ 
\begin{array}{cccc}
-\partial _{0}+\partial _{3} & \partial _{1}-\mathrm{i}\partial _{2} & 
\partial _{y}^{\beta }-\partial _{y}^{\theta } & \partial _{y}^{\zeta
}-i\partial _{y}^{\eta } \\ 
\partial _{1}+\mathrm{i}\partial _{2} & -\partial _{0}-\partial _{3} & 
\partial _{y}^{\zeta }+i\partial _{y}^{\eta } & \partial _{y}^{\beta
}+\partial _{y}^{\theta } \\ 
\partial _{y}^{\beta }-\partial _{y}^{\theta } & \partial _{y}^{\zeta
}-i\partial _{y}^{\eta } & -\partial _{0}-\partial _{3} & -\partial _{1}+%
\mathrm{i}\partial _{2} \\ 
\partial _{y}^{\zeta }+i\partial _{y}^{\eta } & \partial _{y}^{\beta
}+\partial _{y}^{\theta } & -\partial _{1}-\mathrm{i}\partial _{2} & 
-\partial _{0}+\partial _{3}%
\end{array}%
\right]  \\ 
+\mathrm{i}\left[ 
\begin{array}{cccc}
0 & 0 & \partial _{z}^{\beta }+\partial _{z}^{\theta } & \partial
_{z}^{\zeta }+i\partial _{z}^{\eta } \\ 
0 & 0 & \partial _{z}^{\zeta }-i\partial _{z}^{\eta } & \partial _{z}^{\beta
}-\partial _{z}^{\theta } \\ 
-\partial _{z}^{\beta }-\partial _{z}^{\theta } & -\partial _{z}^{\zeta
}-i\partial _{z}^{\eta } & 0 & 0 \\ 
-\partial _{z}^{\zeta }+i\partial _{z}^{\eta } & -\partial _{z}^{\beta
}+\partial _{z}^{\theta } & 0 & 0%
\end{array}%
\right] 
\end{array}
\label{G00}
\end{equation}

\begin{eqnarray}
&&G_{0}\left[ \varphi \right] =-\mathrm{i}\frac{\mathrm{h}}{\mathrm{c}}%
\sum_{w,p_{1},p_{2},p_{3},n^{\beta },s^{\beta },n^{\zeta },s^{\zeta
},n^{\eta },s^{\eta },n^{\theta },s^{\theta }}\check{g}(w,  \nonumber \\
&&p_{1},p_{2},p_{3},n^{\beta },s^{\beta },n^{\zeta },s^{\zeta },n^{\eta
},s^{\eta },n^{\theta },s^{\theta })  \nonumber \\
&&\sum_{k=0}^{3}c_{k}(w,p_{1},p_{2},p_{3},n^{\beta },s^{\beta },n^{\zeta
},s^{\zeta },n^{\eta },s^{\eta },n^{\theta },s^{\theta })\times   \nonumber
\\
&&\times \exp (-\mathrm{i}\frac{\mathrm{h}}{\mathrm{c}}%
(wx_{0}+p_{1}x_{1}+p_{2}x_{2}+p_{3}x_{3}+  \label{fuuu} \\
&&+n^{\beta }y^{\beta }+s^{\beta }z^{\beta }\ +n^{\zeta }y^{\zeta }+s^{\zeta
}z^{\zeta }+  \nonumber \\
&&+n^{\eta }y^{\eta }+s^{\eta }z^{\eta }+n^{\theta }y^{\theta }+s^{\theta
}z^{\theta }))\mbox{.}  \nonumber
\end{eqnarray}

\noindent
here 
\[c_{k}(w,p_{1},p_{2},p_{3},n^{\beta },s^{\beta },n^{\zeta },s^{\zeta
},n^{\eta },s^{\eta },n^{\theta },s^{\theta })\] 
is an eigenvector of \[\check{%
g}(w,p_{1},p_{2},p_{3},n^{\beta },s^{\beta },n^{\zeta },s^{\zeta },n^{\eta
},s^{\eta },n^{\theta },s^{\theta })\]
and

\begin{eqnarray}
&&\check{g}(w,p_{1},p_{2},p_{3},n^{\beta },s^{\beta },n^{\zeta },s^{\zeta
},n^{\eta },s^{\eta },n^{\theta },s^{\theta }) := \label{gu} \\
&&:=\beta ^{\left[ 0\right] }w+\beta ^{\left[ 1\right] }p_{1}+\beta ^{\left[
2\right] }p_{2}+\beta ^{\left[ 3\right] }p_{3}+ \nonumber\\
&&+\gamma ^{\left[ 0\right] }n^{\beta }+\beta ^{\left[ 4\right] }s^{\beta
}-\gamma _{\zeta }^{[0]}n^{\zeta }+\zeta ^{\lbrack 4]}s^{\zeta }-\nonumber \\
&&-\gamma _{\eta }^{[0]}n^{\eta }-\eta ^{\lbrack 4]}s^{\eta }+\gamma
_{\theta }^{[0]}n^{\theta }+\theta ^{\lbrack 4]}s^{\theta }\mbox{.}\nonumber
\end{eqnarray}

Here 
\[\left\{c_{0},c_{1},c_{2},c_{3}\right\} \] 
is an orthonormalized basis of the complex4-vectors space.

Functions 

\begin{eqnarray}
&&c_{k}(w,p_{1},p_{2},p_{3},n^{\beta },s^{\beta },n^{\zeta },s^{\zeta
},n^{\eta },s^{\eta },n^{\theta },s^{\theta })\times   \label{eg} \\
&&\times \exp (-\mathrm{i}\frac{\mathrm{h}}{\mathrm{c}}%
(wx_{0}+p_{1}x_{1}+p_{2}x_{2}+p_{3}x_{3}+  \nonumber \\
&&+n^{\beta }y^{\beta }+s^{\beta }z^{\beta }\ +  \nonumber \\
&&+n^{\zeta }y^{\zeta }+s^{\zeta }z^{\zeta }++n^{\eta }y^{\eta }+s^{\eta
}z^{\eta }+n^{\theta }y^{\theta }+s^{\theta }z^{\theta }))  \nonumber
\end{eqnarray}

\noindent
are eigenvectors of operator $G_{0}$.

\[
\varphi _{y}^{\zeta }:=c(w,\mathbf{p},f)\exp (-\mathrm{i}\frac{\mathrm{h}}{%
\mathrm{c}}(wx_{0}+\mathbf{px}+\gamma _{\zeta }^{[0]}fy^{\zeta })) 
\]

\noindent
is \textit{a red lower chrome function,}

\[
\varphi _{z}^{\zeta }:=c(w,\mathbf{p},f)\exp (-\mathrm{i}\frac{\mathrm{h}}{%
\mathrm{c}}(wx_{0}+\mathbf{px}-\mathrm{i}\zeta ^{\lbrack 4]}fz^{\zeta })) 
\]

\noindent
is \textit{a red upper chrome function,}

\[
\varphi _{y}^{\eta }:=c(w,\mathbf{p},f)\exp (-\mathrm{i}\frac{\mathrm{h}}{%
\mathrm{c}}(wx_{0}+\mathbf{px}+\gamma _{\eta }^{[0]}fy^{\eta })) 
\]

\noindent
is \textit{a green lower chrome function,}

\[
\varphi _{z}^{\eta }:=c(w,\mathbf{p},,f)\exp (-\mathrm{i}\frac{\mathrm{h}}{%
\mathrm{c}}(wx_{0}+\mathbf{px}-\mathrm{i}\eta ^{\lbrack 4]}fz^{\eta })) 
\]

\noindent
is \textit{a green upper chrome function,}

\[
\varphi _{y}^{\theta }:=c(w,\mathbf{p},f)\exp (-\mathrm{i}\frac{\mathrm{h}}{%
\mathrm{c}}(wx_{0}+\mathbf{px+}\gamma _{\theta }^{[0]}fy^{\theta })) 
\]

\noindent
is \textit{a blue lower function,}

\[
\varphi _{z}^{\theta }:=c(w,\mathbf{p},s^{\theta })\exp (-\mathrm{i}\frac{%
\mathrm{h}}{\mathrm{c}}(wx_{0}+\mathbf{px}-\mathrm{i}\theta ^{\lbrack
4]}fz^{\theta })) 
\]

\noindent
is \textit{a blue upper chrome function.}

Operator $-\partial _{y}^{\zeta }\partial _{y}^{\zeta }$ is called \textit{a
red lower chrome operator}, $-\partial _{z}^{\zeta }\partial _{z}^{\zeta }$ is 
\textit{a red upper chrome operator,}

$-\partial _{y}^{\eta }\partial _{y}^{\eta }$ is called \textit{a green lower
chrome operator}, $-\partial _{z}^{\eta }\partial _{z}^{\eta }$ is \textit{a
green upper chrome operator, }

$-\partial _{y}^{\theta }\partial _{y}^{\theta }$ is called \textit{a blue
lower chrome operator}, $-\partial _{z}^{\theta }\partial _{z}^{\theta }$ is 
\textit{a blue upper chrome operator}

For example, if $\varphi _{z}^{\zeta }$ is a red upper chrome function then

\begin{eqnarray*}
-\partial _{y}^{\zeta }\partial _{y}^{\zeta }\varphi _{z}^{\zeta }
&=&-\partial _{y}^{\eta }\partial _{y}^{\eta }\varphi _{z}^{\zeta
}=-\partial _{z}^{\eta }\partial _{z}^{\eta }\varphi _{z}^{\zeta }= \\
&=&-\partial _{y}^{\theta }\partial _{y}^{\theta }\varphi _{z}^{\zeta
}=-\partial _{z}^{\theta }\partial _{z}^{\theta }\varphi _{z}^{\zeta }=0
\end{eqnarray*}
but 
\[
-\partial _{z}^{\zeta }\partial _{z}^{\zeta }\varphi _{z}^{\zeta }=-\left( 
\frac{\mathrm{h}}{\mathrm{c}}f\right) ^{2}\varphi _{z}^{\zeta }\mbox{.}
\]

Because
\[
G_{0}\left[ \varphi \right] =0                 
\]
then
\[
UG_{0}U^{-1}U\left[ \varphi \right] =0 
\]

If $U=U_{1,2}\left( \alpha \right) $ then $G_{0}\rightarrow U_{1,2}\left(
\alpha \right) G_{0}U_{1,2}^{-1}\left( \alpha \right) $ and $\left[ \varphi %
\right] \rightarrow U_{1,2}\left( \alpha \right) \left[ \varphi \right] $.

In this case:

$\partial _{1}\rightarrow \partial _{1}^{\prime }:=\allowbreak \left( \cos
\alpha \cdot \partial _{1}-\sin \alpha \cdot \partial _{2}\right) $,

$\partial _{2}\rightarrow \partial _{2}^{\prime }:=\left( \cos \alpha \cdot
\partial _{2}+\sin \alpha \cdot \partial _{1}\right) $,

$\partial _{0}\rightarrow \partial _{0}^{\prime }:=\partial _{0}$,

$\partial _{3}\rightarrow \partial _{3}^{\prime }:=\partial _{3}$,

$\partial _{y}^{\beta }\rightarrow \partial _{y}^{\beta \prime }:=\partial
_{y}^{\beta }$,

$\partial _{z}^{\beta }\rightarrow \partial _{z}^{\beta \prime }:=\partial
_{z}^{\beta }$,

$\partial _{y}^{\zeta }\rightarrow \partial _{y}^{\zeta \prime }:=\left(
\cos \alpha \cdot \partial _{y}^{\zeta }-\sin \alpha \cdot \partial
_{y}^{\eta }\right) $,

$\partial _{y}^{\eta }\rightarrow \partial _{y}^{\eta \prime }:=\left( \cos
\alpha \cdot \partial _{y}^{\eta }+\sin \alpha \cdot \partial _{y}^{\zeta
}\right) $,

$\partial _{z}^{\zeta }\rightarrow \partial _{z}^{\zeta \prime }:=\left(
\cos \alpha \cdot \partial _{z}^{\zeta }+\allowbreak \sin \alpha \cdot
\partial _{z}^{\eta }\right) $,

$\partial _{z}^{\eta }\rightarrow \partial _{z}^{\eta \prime }:=\left( \cos
\alpha \cdot \partial _{z}^{\eta }-\sin \alpha \cdot \partial _{z}^{\zeta
}\right) $,

$\partial _{y}^{\theta }\rightarrow \partial _{y}^{\theta \prime }:=\partial
_{y}^{\theta }$,

$\partial _{z}^{\theta }\rightarrow \partial _{z}^{\theta \prime }:=\partial
_{z}^{\theta }$.

Therefore, 
\begin{eqnarray*}
-\partial _{z}^{\zeta \prime }\partial _{z}^{\zeta \prime }\varphi
_{z}^{\zeta } &=&\left( f\frac{\mathrm{h}}{\mathrm{c}}\cos \alpha \right)
^{2}\cdot \varphi _{z}^{\zeta }\mbox{,} \\
-\partial _{z}^{\eta \prime }\partial _{z}^{\eta \prime }\varphi _{z}^{\zeta
} &=&\left( -\sin \alpha \cdot f\frac{\mathrm{h}}{\mathrm{c}}\right)
^{2}\varphi _{z}^{\zeta }\mbox{.}
\end{eqnarray*}

If $\alpha =-\frac{\pi }{2}$ then 
\begin{eqnarray*}
-\partial _{z}^{\zeta \prime }\partial _{z}^{\zeta \prime }\varphi
_{z}^{\zeta } &=&0\mbox{,} \\
-\partial _{z}^{\eta \prime }\partial _{z}^{\eta \prime }\varphi _{z}^{\zeta
} &=&\left( f\frac{\mathrm{h}}{\mathrm{c}}\right) ^{2}\varphi _{z}^{\zeta }%
\mbox{.}
\end{eqnarray*}

That is under such rotation the red state becomes the green state.

If $U=U_{3,2}\left( \alpha \right) $ then $G_{0}\rightarrow U_{3,2}\left(
\alpha \right) G_{0}U_{3,2}^{-1}\left( \alpha \right) $ and $\left[ \varphi %
\right] \rightarrow U_{3,2}\left( \alpha \right) \left[ \varphi \right] $.

In this case:

$\partial _{0}\rightarrow \partial _{0}^{\prime }:=\partial _{0}$,

$\partial _{1}\rightarrow \partial _{1}^{\prime }:=\partial _{1}$,

$\partial _{2}\rightarrow \partial _{2}^{\prime }:=\left( \cos \alpha \cdot
\partial _{2}+\sin \alpha \cdot \partial _{3}\right) $,

$\partial _{3}\rightarrow \partial _{3}^{\prime }:=\left( \cos \alpha \cdot
\partial _{3}-\sin \alpha \cdot \partial _{2}\right) $,

$\partial _{y}^{\beta }\rightarrow \partial _{y}^{\beta \prime }:=\partial
_{y}^{\beta }$,

$\partial _{y}^{\zeta }\rightarrow \partial _{y}^{\zeta \prime }:=\partial
_{y}^{\zeta }$,

$\partial _{y}^{\eta }\rightarrow \partial _{y}^{\eta \prime }:=\left( \cos
\alpha \cdot \partial _{y}^{\eta }-\sin \alpha \cdot \partial _{y}^{\theta
}\right) $,

$\partial _{y}^{\theta }\rightarrow \partial _{y}^{\theta \prime }:=\left(
\cos \alpha \cdot \partial _{y}^{\theta }+\sin \alpha \cdot \partial
_{y}^{\eta }\right) $,

$\partial _{z}^{\beta }\rightarrow \partial _{z}^{\beta \prime }:=\partial
_{z}^{\beta }$,

$\partial _{z}^{\zeta }\rightarrow \partial _{z}^{\zeta \prime }:=\partial
_{z}^{\zeta }$,

$\partial _{z}^{\eta }\rightarrow \partial _{z}^{\eta \prime }:=\left( \cos
\alpha \cdot \partial _{z}^{\eta }-\sin \alpha \cdot \partial _{z}^{\theta
}\right) $,

$\partial _{z}^{\theta }\rightarrow \partial _{z}^{\theta \prime }:=\left(
\cos \alpha \cdot \partial _{z}^{\theta }+\sin \alpha \cdot \partial
_{z}^{\eta }\right) $,

Therefore, if $\varphi _{y}^{\eta }$ is a green lower chrome function then

\begin{eqnarray*}
-\partial _{z}^{\eta \prime }\partial _{z}^{\eta \prime }\varphi _{y}^{\eta
} &=&\left( \frac{\mathrm{h}}{\mathrm{c}}\cos \alpha \cdot f\right)
^{2}\cdot \varphi _{y}^{\eta }\mbox{,} \\
-\partial _{y}^{\theta \prime }\partial _{y}^{\theta \prime }\varphi
_{y}^{\eta } &=&\left( \frac{\mathrm{h}}{\mathrm{c}}\sin \alpha \cdot
f\right) ^{2}\cdot \varphi _{y}^{\eta }\mbox{.}
\end{eqnarray*}

If $\alpha =\pi /2$ then

\begin{eqnarray*}
-\partial _{z}^{\eta \prime }\partial _{z}^{\eta \prime }\varphi _{y}^{\eta
} &=&0\mbox{,} \\
-\partial _{y}^{\theta \prime }\partial _{y}^{\theta \prime }\varphi
_{y}^{\eta } &=&\left( \frac{\mathrm{h}}{\mathrm{c}}f\right) ^{2}\cdot
\varphi _{y}^{\eta }\mbox{.}
\end{eqnarray*}

That is under such rotation the green state becomes blue state.

If $U=U_{3,1}\left( \alpha \right) $ then $G_{0}\rightarrow U_{3,1}\left(
\alpha \right) G_{0}U_{3,1}^{-1}\left( \alpha \right) $ and $\left[ \varphi %
\right] \rightarrow U_{3,1}\left( \alpha \right) \left[ \varphi \right] $.

In this case:

$\partial _{0}\rightarrow \partial _{0}^{\prime }:=\partial _{0}$,

$\partial _{1}\rightarrow \partial _{1}^{\prime }:=\left( \cos \alpha \cdot
\partial _{1}-\sin \alpha \cdot \partial _{3}\right) $,

$\partial _{2}\rightarrow \partial _{2}^{\prime }:=\partial _{2}$,

$\partial _{3}\rightarrow \partial _{3}^{\prime }:=\left( \cos \alpha \cdot
\partial _{3}+\sin \alpha \cdot \partial _{1}\right) $,

$\partial _{y}^{\beta }\rightarrow \partial _{3}^{\prime }:=\partial
_{y}^{\beta }$,

$\partial _{y}^{\zeta }\rightarrow \partial _{y}^{\zeta \prime }:=\left(
\cos \alpha \cdot \partial _{y}^{\zeta }+\sin \alpha \cdot \partial
_{y}^{\theta }\right) $,

$\partial _{y}^{\eta }\rightarrow \partial _{y}^{\eta \prime }:=\partial
_{y}^{\eta }$,

$\partial _{y}^{\theta }\rightarrow \partial _{y}^{\theta \prime }:=\left(
\cos \alpha \cdot \partial _{y}^{\theta }-\sin \alpha \cdot \partial
_{y}^{\zeta }\right) $,

$\partial _{z}^{\beta }\rightarrow \partial _{z}^{\beta \prime }:=\partial
_{z}^{\beta }$,

$\partial _{z}^{\zeta }\rightarrow \partial _{z}^{\zeta \prime }:=\left(
\cos \alpha \cdot \partial _{z}^{\zeta }-\sin \alpha \cdot \partial
_{z}^{\theta }\right) $,

$\partial _{z}^{\eta }\rightarrow \partial _{z}^{\eta \prime }:=\partial
_{z}^{\eta }$,

$\partial _{z}^{\theta }\rightarrow \partial _{z}^{\theta \prime }:=\left(
\cos \alpha \cdot \partial _{z}^{\theta }+\sin \alpha \cdot \partial
_{z}^{\zeta }\right) $.

Therefore, 
\begin{eqnarray*}
-\partial _{z}^{\zeta \prime }\partial _{z}^{\zeta \prime }\varphi
_{z}^{\zeta } &=&-\left( f\frac{\mathrm{h}}{\mathrm{c}}\cos \alpha \right)
^{2}\cdot \varphi _{z}^{\zeta }\mbox{,} \\
-\partial _{z}^{\theta \prime }\partial _{z}^{\theta \prime }\varphi
_{z}^{\zeta } &=&-\left( \sin \alpha \cdot f\frac{\mathrm{h}}{\mathrm{c}}%
\right) ^{2}\varphi _{z}^{\zeta }\mbox{.}
\end{eqnarray*}

If $\alpha =\pi /2$ then 
\begin{eqnarray*}
-\partial _{z}^{\zeta \prime }\partial _{z}^{\zeta \prime }\varphi
_{z}^{\zeta } &=&0\mbox{,} \\
-\partial _{z}^{\theta \prime }\partial _{z}^{\theta \prime }\varphi
_{z}^{\zeta } &=&-\left( f\frac{\mathrm{h}}{\mathrm{c}}\right) ^{2}\varphi
_{z}^{\zeta }\mbox{.}
\end{eqnarray*}

That is under such rotation the red state becomes the blue state. Thus at the Cartesian turns chrome of a state is changed.

One of ways of elimination of this noninvariancy consists in the following. Calculations \ref{acs} give the grounds to assume that 
some oscillations of quarks states 
bend time-space in such a way that acceleration of the bent system in relation to initial system submits 
to the following law (Figure 3):

\begin{figure}[htbp]
\centering
\includegraphics[width=.75\textwidth]{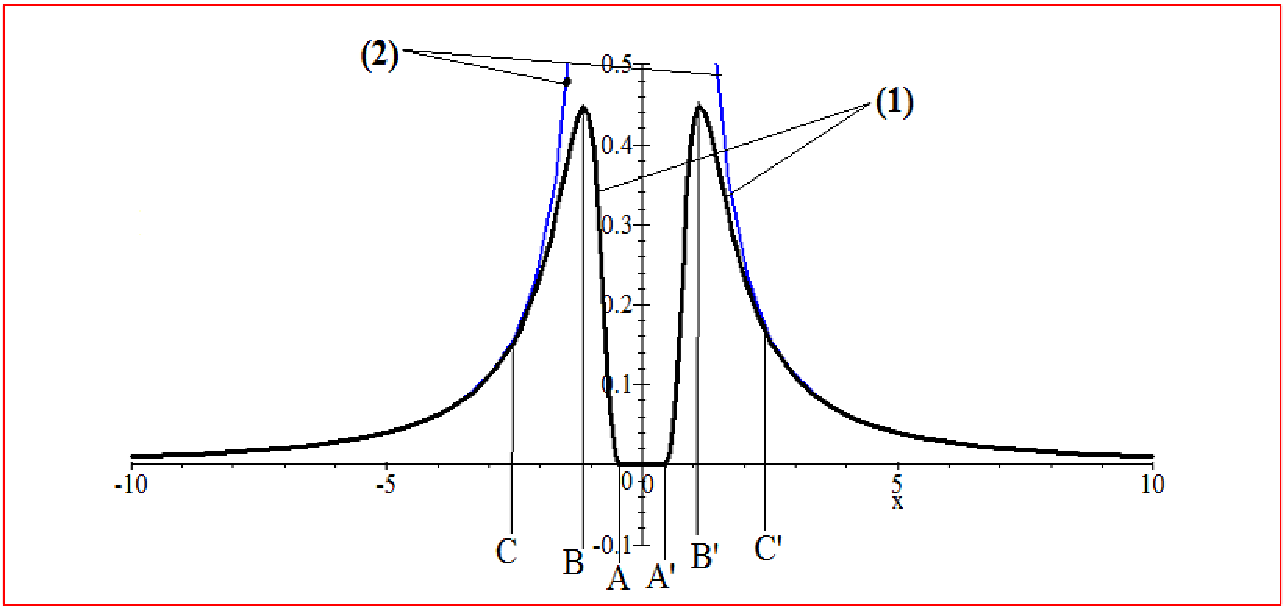}
\caption{acceleration of the bent system in relation to initial system. (see Figure 2).
The acceleration plot is line (1) and the  line (2) is plot of $\lambda /\mathbf{x}^{2}$.}
\end{figure}

\[
g\left( t,\mathbf{x}\right) =\mathrm{c}\lambda /\left( \mathbf{x}^{2}\cosh
^{2}\left( \lambda t/\mathbf{x}^{2}\right) \right) \mbox{.}
\]

Here the acceleration plot is line (1) and the  line (2) is plot of $\lambda /\mathbf{x}^{2}$.

Hence, to the right from point $C^{\prime } $ and to the left from poin $C$ t%
\textit{he Newtonian gravitation law} is carried out.

$AA^{\prime } $ is \textit{the Asymptotic Freedom Zone}.

$CB$ and $B^{\prime } C^{\prime } $ is \textit{the Confinement Zone.}

Let in the potential hole $AA^{\prime } $ there are three quarks $\varphi
_{y}^{\zeta }$, $\varphi _{y}^{\eta }$, $\varphi _{y}^{\theta }$. Their
general state function is determinant with elements of the following type: $%
\varphi _{y}^{\zeta \eta \theta }:=\varphi _{y}^{\zeta }\varphi _{y}^{\eta
}\varphi _{y}^{\theta }$. In this case:

\[
-\partial _{y}^{\zeta }\partial _{y}^{\zeta }\varphi _{y}^{\zeta \eta \theta
}=\left( \frac{\mathrm{h}}{\mathrm{c}}f\right) ^{2}\varphi _{y}^{\zeta \eta
\theta }
\]

\noindent
and under rotation $U_{1,2}\left( \alpha \right) $:

\begin{eqnarray*}
-\partial _{y}^{\zeta \prime }\partial _{y}^{\zeta \prime }\varphi
_{y}^{\zeta \eta \theta } &=& \\
&=&\left( \frac{\mathrm{h}}{\mathrm{c}}f\right) ^{2}\left( \gamma _{\zeta
}^{[0]}\cos \alpha -\gamma _{\eta }^{[0]}\sin \alpha \right) ^{2}\left(
\varphi _{y}^{\zeta }\varphi _{y}^{\eta }\varphi _{y}^{\theta }\right) = \\
&=&\left( \frac{\mathrm{h}}{\mathrm{c}}f\right) ^{2}\varphi _{y}^{\zeta \eta
\theta }\mbox{.}
\end{eqnarray*}

That is at such turns the quantity of red chrome remains. 

As and for all other Cartesian turns and for all other chromes.

Baryons $\Delta ^{-}=ddd$, $\Delta ^{++}=uuu$, $\Omega ^{-}=sss$ belong to
such structures.

If $U=U_{1,0}\left( \alpha \right) $ then $G_{0}\rightarrow
U_{1,0}^{-1\ddagger }\left( \alpha \right) G_{0}U_{1,0}^{-1}\left( \alpha
\right) $ and $\left[ \varphi \right] \rightarrow U_{1,0}\left( \alpha
\right) \left[ \varphi \right] $.

In this case:

$\partial _{0}\rightarrow \partial _{0}^{\prime }:=\left( \cosh \alpha \cdot
\partial _{0}+\sinh \alpha \cdot \partial _{1}\right) $,

$\partial _{1}\rightarrow \partial _{1}^{\prime }:=\left( \cosh \alpha \cdot
\partial _{1}+\sinh \alpha \cdot \partial _{0}\right) $,

$\partial _{2}\rightarrow \partial _{2}^{\prime }:=\partial _{2}$,

$\partial _{3}\rightarrow \partial _{3}^{\prime }:=\partial _{3}$,

$\partial _{y}^{\beta }\rightarrow \partial _{y}^{\beta \prime }:=\partial
_{y}^{\beta }$,

$\partial _{y}^{\zeta }\rightarrow \partial _{y}^{\zeta \prime }:=\partial
_{y}^{\zeta }$,

$\partial _{y}^{\eta }\rightarrow \partial _{y}^{\eta \prime }:=\left( \cosh
\alpha \cdot \partial _{y}^{\eta }-\sinh \alpha \cdot \partial _{z}^{\theta
}\right) $,

$\partial _{y}^{\theta }\rightarrow \partial _{y}^{\theta \prime }:=\left(
\cosh \alpha \cdot \partial _{y}^{\theta }+\sinh \alpha \cdot \partial
_{z}^{\eta }\right) $,

$\partial _{z}^{\beta }\rightarrow \partial _{z}^{\beta \prime }:=\partial
_{z}^{\beta }$,

$\partial _{z}^{\zeta }\rightarrow \partial _{z}^{\zeta \prime }:=\partial
_{z}^{\zeta }$,

$\partial _{z}^{\eta }\rightarrow \partial _{z}^{\eta \prime }:=\left( \cosh
\alpha \cdot \partial _{z}^{\eta }+\sinh \alpha \cdot \partial _{y}^{\theta
}\right) $,

$\partial _{z}^{\theta }\rightarrow \partial _{z}^{\theta \prime }:=\left(
\cosh \alpha \cdot \partial _{z}^{\theta }-\sinh \alpha \cdot \partial
_{y}^{\eta }\right) $.

Therefore,

\begin{eqnarray*}
-\partial _{y}^{\eta \prime }\partial _{y}^{\eta \prime }\varphi _{y}^{\eta
} &=&\left( 1+\sinh ^{2}\alpha \right) \cdot \left( \frac{\mathrm{h}}{%
\mathrm{c}}f\right) ^{2}\varphi _{y}^{\eta }\mbox{,} \\
-\partial _{z}^{\theta \prime }\partial _{z}^{\theta \prime }\varphi
_{y}^{\eta } &=&\sinh ^{2}\alpha \cdot \left( \frac{\mathrm{h}}{\mathrm{c}}%
f\right) ^{2}\varphi _{y}^{\eta }\mbox{.}
\end{eqnarray*}

Similarly chromes and grades change for other states and under other Lorentz
transformation.

One of ways of elimination of this noninvariancy is the following:

Let
\[
\varphi _{yz}^{\zeta \eta \theta }:=\varphi _{y}^{\zeta }\varphi _{y}^{\eta
}\varphi _{y}^{\theta }\varphi _{z}^{\zeta }\varphi _{z}^{\eta }\varphi
_{z}^{\theta }\mbox{.} 
\]

Under transformation $U_{1,0}\left( \alpha \right) $:
\[
-\partial _{z}^{\theta \prime }\partial _{z}^{\theta \prime }\varphi
_{yz}^{\zeta \eta \theta }=-\left( \mathrm{i}\frac{\mathrm{h}}{\mathrm{c}}%
f\right) ^{2}\varphi _{yz}^{\zeta \eta \theta }. 
\]

That is a magnitude of red chrome of this state doesn't depend on angle $%
\alpha $.

This condition is satisfied for all chromes and under all Lorentz's
transformations.

Pairs of baryons 
\begin{eqnarray*}
&&\left\{ p=uud,n=ddu\right\} , \\
&&\left\{ \Sigma ^{+}=uus,\Xi ^{0}=uss\right\} , \\
&&\left\{ \Delta ^{+}=uud,\Delta ^{0}=udd\right\} 
\end{eqnarray*}
belong to such structures.

Baryons represent one of ways of elimination of the chrome noninvariancy under Cartesian's and under Lorentz's transformations.

\section{Conclusion}

The Quark Theory is a part uf the Probability Theory


\begin{thebibliography}{99}

\bibitem{Q1}
Gunn Quznetsov, {\it Logical foundation of fundamental theoretical physics}, 
LAMBERT Academic Publishing (2013).

\bibitem{Q2}
Gunn Quznetsov, {\it Probabilistic Treatment of Gauge Theories},
Nova Science Pub Inc, (2007).

\bibitem{Md}
Madelung, E., {\it Die Mathematischen Hilfsmittel des Physikers} 
Springer Verlag, (1957) p.29

\bibitem{Z} For instance, Ziman J. M. {\it Elements of Advanced Quantum Theory}, Cambridge (1969) p.32.

\bibitem{peak} For instance, Peak L. and Varvell, K., {\it The Physics of The Standard Model}, 
part 2. (2002), p.37

\bibitem{Pf} Waclaw Sierpinski. {\it Pythagorean Triangles}. — Dover, 2003. — ISBN 978-0-486-43278-6.

\bibitem{psk}  For instance, Peskin M.\, E., Schroeder D.\, V. {\it An Introduction
to Quantum Field Theory}, Perseus Books Publishing, L.L.C., 1995.

\bibitem{Kane2} Gordon Kane, {\it Modern Elementary Particle Physics}, Addison-Wesley Publ. Comp., (1993), p.93

\bibitem{DVB} Dvoeglazov, V. V., Additional Equations Derived from the Ryder
Postulates in the (1/2,0)+(0,1/2) Representation of the Lorentz Group.
hep-th/9906083. \textit{Int. J. Theor. Phys. }{\bf  37} (1998) 1909. \textit{Helv. Phys. Acta }
{\bf  70} (1997) 677. \textit{Fizika B }{\bf  6} (1997) 75; \textit{Int. J. Theor. Phys.} {\bf  34%
} (1995) 2467.\textit{ Nuovo Cimento} {\bf  108} A (1995) 1467. \textit{Nuovo Cimento} {\bf  111}
B (1996) 483. \textit{Int. J. Theor. Phys.} {\bf  36} (1997) 635.

\bibitem{AV} Ahluwalia, D. V.,  (j,0)+(0,j) Covariant spinors and causal
propagators based on Weinberg formalism. nucl-th/9905047. \textit{Int. J. Mod. Phys.}
\textit{A }{\bf  11} (1996) 1855.

\bibitem{Kane} For instance, Kane, G. {\it Modern Elementary Particle Physics},
Addison-Wesley Publishing Company, Inc. (1987), formulas (7.9), (7.10), (7.18) ) 

\bibitem{O1} \{M. L. Cantor, A. S. Solodovnikov, \textit{Hipercomplex numbers%
}, Moscow, (1973), p.99\}, Kantor, I. L.; Solodownikow, A. S. (1978), \textit{Hyperkomplexe Zahlen}, Leipzig: B.G. Teubner

\bibitem{O2} O. V. Mel'nikov, V. N. Remeslennikov, et el., \textit{General
Algebra}, Moscow, (1990), p.396

\bibitem{ZH} V. A. Zhelnorovich, {\it Theory of spinors. Application to mathematics and physics}, 
Moscow,  (1982), p.21.

\bibitem{Miller} Miller, Chris. Cosmic Hide and Seek: the Search for the Missing Mass.\\ 
http://www.eclipse.net/~cmmiller/DM/. And see Van Den Bergh, Sidney. \textit{The early history of Dark Matter}. 
preprint, astro-ph/9904251.

\bibitem{Hawley} Hawley, J.F. and K.A. Holcomb. \textit{Foundations of modem cosmology}. Oxford University Press, New York, 1998.

\bibitem{Begeman} K. G. Begeman, A. H. Broeils and R. H. Sanders, 1991, MNRAS, 249, 523

\bibitem{HB} Peter Coles, ed., {\it Routledge Critical Dictionary of the New Cosmology}. Routledge, 2001,  202.




\end{thebibliography}
\end{document}